%% file: esqpt.tex
\documentclass[seceqn]{elsart}
\journal{Annals of Physics (N.Y.)}

\usepackage{graphicx}
\usepackage{citesort}
\usepackage{stackrel}
\usepackage{amsmath}
\usepackage{amssymb}
\newcommand{\<}[1]{\hspace{-0.11111em}#1\hspace{-0.11111em}}
\newcommand{\rtrim}[1]{#1\hspace{-0.11111em}}
\DeclareRobustCommand{\abs}[1]{\lvert#1\rvert}

\DeclareRobustCommand{\negThinspace}{\hspace{-.0833em}}

\DeclareRobustCommand{\grp}[1]{\mathrm{#1}}
\DeclareRobustCommand{\grpu}[2][]{\grp{U}_{#1}\negThinspace(#2)}
\DeclareRobustCommand{\grpsu}[2][]{\grp{SU}_{#1}\negThinspace(#2)}
\DeclareRobustCommand{\grpsp}[2][]{\grp{Sp}_{#1}\negThinspace(#2)}

\DeclareRobustCommand{\grpso}[2][]{\grp{SO}_{#1}\negThinspace(#2)}

\DeclareRobustCommand{\Nhat}{\hat{N}}
\DeclareRobustCommand{\Shat}{\hat{S}}
\DeclareRobustCommand{\Svechat}{\mathrm{\mathbf{\hat{S}}}}

\DeclareRobustCommand{\lhat}{\hat{l}}
\DeclareRobustCommand{\Dhat}{\hat{D}}

\DeclareRobustCommand{\Hhat}{\hat{H}}

\DeclareRobustCommand{\xic}{{\xi_\mathrm{c}}}
\DeclareRobustCommand{\xicex}{{\xi_\mathrm{c}^\mathrm{ex}}}
\DeclareRobustCommand{\gqn}[2]{\stackrel[#2]{}{#1}}
\DeclareRobustCommand{\Ec}{{E_\mathrm{c}}}
\DeclareRobustCommand{\kc}{{k_\mathrm{c}}}

\newcommand{\hbarfracinline}{{\hbar^2/(2m)}}

\begin{document}

\begin{frontmatter}

\title{Excited state quantum phase transitions in many-body systems}

\author[yale]{M.~A.~Caprio\corauthref{cor1}},
\author[charles,ect]{P.~Cejnar}, and
\author[yale]{F.~Iachello}

\address[yale]{Center for Theoretical Physics, Sloane Physics Laboratory, 
Yale University, New Haven, Connecticut 06520-8120, USA}
\address[charles]{Institute of Particle and Nuclear Physics, 
Faculty of Mathematics and Physics, Charles University,
V Hole\v{s}ovi\v{c}k\'{a}ch 2, 180\protect\,00 Praha, Czech Republic}
\address[ect]{European Centre for Theoretical
Studies in Nuclear Physics and Related Areas,
Strada delle Tabarelle 286, 38050 Villazzano (Trento), Italy}
\corauth[cor1]{Corresponding author.}

\begin{abstract}
Phenomena analogous to ground state quantum phase transitions have
recently been noted to occur among states throughout the excitation
spectra of certain many-body models.  These excited state phase
transitions are manifested as simultaneous singularities in the
eigenvalue spectrum (including the gap or level density), order
parameters, and wave function properties.  In this article, the
characteristics of excited state quantum phase transitions are
investigated.  The finite-size scaling behavior is determined at the
mean field level.  It is found that excited state quantum phase
transitions are universal to two-level bosonic and fermionic models
with pairing interactions.
\end{abstract}

% suppress "Key words:" heading
\makeatletter
\def\@keywordheading{}
\makeatother  

\begin{keyword}
\PACS 03.65.Fd \sep 03.65.Sq \sep 64.60.-i
\end{keyword}
% X 02.20.Qs -- General: Math methods: Group: General properties, structure, repn. of Lie
% 03.65.Fd -- General: Quantum mechanics: Algebraic methods
% 03.65.Sq -- General: Quantum mechanics: Semiclassical theories and applications
% X 05.70.Fh -- Stat phys, thermo, nonlinear: Thermo: Phase transitions general
% 64.60.-i -- Condensed matter: EOS, phase: Phase transitions general

\end{frontmatter}

%%%%%%%%%%%%%%%%%%%%%%%%%%%%%%%%%%%%%%%%%%%%%%%%%%%%%%%%%%%%%%%%
%%%%%%%%%%%%%%%%%%%%%%%%%%%%%%%%%%%%%%%%%%%%%%%%%%%%%%%%%%%%%%%%

\section{Introduction}
\label{secintro}

Quantum phase transitions (QPTs), or singularities in the evolution of
the ground state properties of a system as a Hamiltonian parameter is
varied, have been extensively studied for various many-body 
systems (\textit{e.g.},
Refs.~\cite{gilmore1978:lipkin,feng1981:ibm-phase,vojta2003:qpt}).
Recently, analogous singular behavior has been noted for states
throughout the excitation spectrum of certain many-body
models~\cite{heiss2002:qpt-instability,heiss2005:lipkin-exceptional,leyvraz2005:lipkin-scaling,heinze2006:ibm-o6-u5-part1-level-dynamics,macek2006:ibm-o6-u5-part2-classical-trajectories,cejnar2006:excited-qpt,heiss2006:lipkin-thermo},
namely the Lipkin model~\cite{lipkin1965:lipkin1} and the interacting
boson model (IBM) for nuclei~\cite{iachello1987:ibm}.  These
singularities have been loosely described as ``excited state quantum
phase transitions'' (ESQPTs)~\cite{cejnar2006:excited-qpt}.  In this
article, we more closely and systematically examine the
characteristics of such excited state singularities as phase
transitions, to provide a foundation for future investigations.  It is
found that excited state quantum phase transitions occur in a much
broader class of many-body models than previously identified.

Ground state QPTs are characterized by a few distinct but related
properties.  The QPT occurs as a ``control parameter'' $\xi$,
controlling an interaction strength in the system's Hamiltonian
$\Hhat(\xi)$, is varied, at some critical value $\xi\<=\xic$.  For
specificity, we take the Hamiltonian to have the conventional form
$\Hhat(\xi)=(1-\xi) \Hhat_1 + \xi \Hhat_2$.  At the critical value:
(1) The ground state energy $E_0$ is nonanalytic as a function of the
control parameter at $\xi\<=\xic$.  (2) The ground state wave function
properties, expressed via ``order parameters'' such as the ground
state expectation values $\langle \Hhat_1
\rangle_0$ or $\langle \Hhat_2 \rangle_0$, are nonanalytic at $\xi\<=\xic$. 
These two properties are not independent, since the evolution of the
ground state energy and that of the order parameters are directly
related by the Feynman-Hellmann
theorem~\cite{feynman1939:molecular-hypervirial}, which gives
$dE_0/d\xi\<=\langle \Hhat_2 \rangle_0 -\langle \Hhat_1 \rangle_0$.  (3) The
gap $\Delta$ between the ground state and the first excited state
vanishes at $\xi\<=\xic$.  (Here we consider only \textit{continuous}
phase transitions.  More specifically, the systems considered in this
article undergo second-order phase transitions, in which discontinuity
occurs in the second derivative of the ground state energy and the
first derivatives of the order parameters.)
Singularities strictly only occur for an infinite number of particles
in the many-body system, but precursors can be observed even for very
modest numbers of particles.  For finite particle number $N$, the
defining characteristic of the QPT is therefore not the presence of a
true singularity but rather well-defined scaling behavior of the
relevant quantities towards their singular large-$N$
limits~\cite{fisher1972:finite-size-scaling}.

\begin{figure}
\begin{center}
\includegraphics*[width=0.7\hsize]{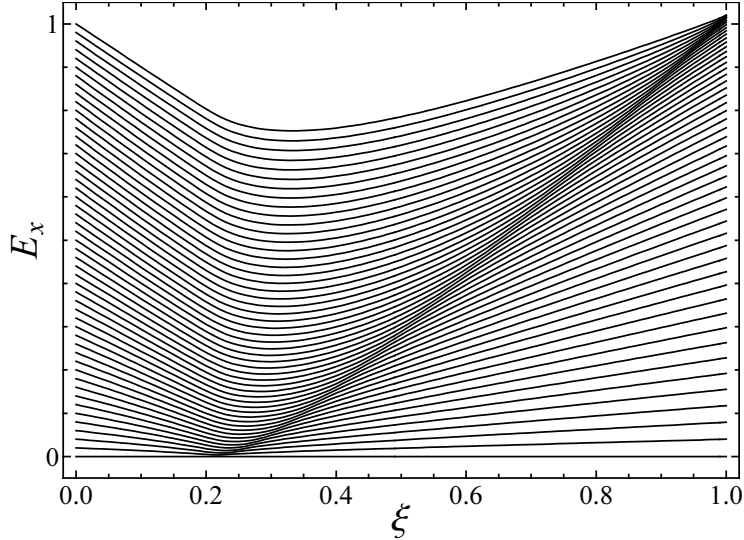}
\end{center}
\caption{Excitation energies for the two-level fermionic pairing
model~(\ref{eqnHPPxi}) with particle number $N\<=100$, at half filling
and zero seniority, as a function of the control parameter $\xi$.}
\label{fig-evoln-pairing}
\end{figure}

For the systems which exhibit \textit{excited state} QPTs, the
vanishing gap between the ground state and first excited state at the
ground state QPT does not occur in isolation.  Rather, there is a
bunching of levels near the ground state, that is, a vanishing of the
average level spacing $\bar{\Delta}$ or an infinite local level
density $\rho\equiv
\bar{\Delta}^{-1}$.  The infinite level density, moreover, propagates
to higher excitation energy (as illustrated for a two-level fermionic
pairing model in Fig.~\ref{fig-evoln-pairing}) as the order parameter
is varied from $\xic$, hence the concept of a continuation of the QPT
to excited states.  The singular level density occurs simultaneously
with singularities in other properties of the excited states $\lvert k
\rangle$, such as the level energy $E_k$ and the expectation values
$\langle \Hhat_1 \rangle_k$ and $\langle \Hhat_2 \rangle_k$.

First, we review the essential properties of the two-level pairing
many-body models, for both bosonic and fermionic constituents
(Sec.~\ref{secmodels}).  We find that ESQPTs are universal to these
models, suggesting that the ESQPT phenomena may be broadly relevant,
at least to systems dominated by pairing interactions.  The
semiclassical analysis of a ``sombrero'' potential provides a basis
for understanding many of the properties of the quantum many-body
ESQPT~\cite{leyvraz2005:lipkin-scaling,cejnar2006:excited-qpt}.  The
semiclassical analysis of
Refs.~\cite{leyvraz2005:lipkin-scaling,cejnar2006:excited-qpt} is
extended in Sec.~\ref{secclass} to address several properties relevant
to the definition of phase transitions.  In particular, the
singularity in the eigenvalue spectrum and the finite-size scaling
behavior for the ESQPT are determined at the mean field level.
Numerical calculations for the full quantum problem are considered in
Sec.~\ref{secobs}, where we investigate manifestations of the ESQPT in
the excitation spectrum and in the properties of ``order parameters''
for the excited states.  Finally, we consider the ESQPT as a boundary
between qualitatively distinct ``phases'' (Sec.~\ref{secphase}).  The
relationship between the $\grpu{n+1}$ two-level boson models and the
two-level pairing model is established for arbitrary dimension in the
appendices, where some further mathematical definitions and
identities are also provided for reference.

%%%%%%%%%%%%%%%%%%%%%%%%%%%%%%%%%%%%%%%%%%%%%%%%%%%%%%%%%%%%%%%%
%%%%%%%%%%%%%%%%%%%%%%%%%%%%%%%%%%%%%%%%%%%%%%%%%%%%%%%%%%%%%%%%

\section{Bosonic and fermionic two-level models}
\label{secmodels}

\begin{figure}
\begin{center}
\includegraphics*[width=0.7\hsize]{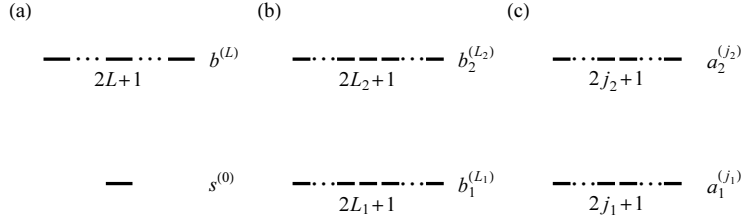}
\end{center}
\caption{
Single-particle level degeneracies for the various
classes of two-level models considered: (a)~the $s$-$b$ boson models,
(b)~the more general two-level bosonic pairing models, and (c)~the
two-level fermionic pairing models.  }
\label{fig-levels}
\end{figure}

Ground state QPTs have been studied extensively (\textit{e.g.},
Refs.~\cite{dieperink1980:ibm-classical,dieperink1980:ibm-phase,feng1981:ibm-phase,vanroosmalen1982:diss})
for the two-level boson models, or $s$-$b$ models, defined in terms of
a singlet boson $s^{(0)}$ and a $(2L+1)$-fold degenerate boson
$b^{(L)}$ [Fig.~\ref{fig-levels}(a)].  Models in this class include
the $\grpu{6}$ interacting boson model (IBM) for nuclei
($L\<=2$)~\cite{iachello1987:ibm}, which is defined in terms of
$s^{(0)}$ and $d^{(2)}$ bosons, and the $\grpu{4}$ vibron model for
molecules ($L\<=1$)~\cite{iachello1995:vibron}.  Also, the Lipkin
model~\cite{lipkin1965:lipkin1} has several isomorphic realizations,
defined variously in terms of systems of interacting fermions,
interacting spins, or interacting bosons (Schwinger realization).
This last realization falls into the two-level boson model
categorization, as the $L\<=0$ case.  So far, excited state QPTs have
been considered in the Lipkin
model~\cite{heiss2005:lipkin-exceptional,leyvraz2005:lipkin-scaling}
and the
IBM~\cite{heinze2006:ibm-o6-u5-part1-level-dynamics,macek2006:ibm-o6-u5-part2-classical-trajectories,cejnar2006:excited-qpt},
both of which are examples of $s$-$b$ two-level models.

The $s$-$b$ two-level models are described by the $\grpu{n+1}$ algebraic structure
\begin{equation}
\label{eqnchainuso}
\grpu{n+1}
\supset
\left\lbrace
\begin{array}{c}
\grpso{n+1}\\
\grpu{n}
\end{array}
\right\rbrace
\supset
\grpso{n}
\supset
\grpso{3},
\end{equation}
where $n\<=2L+1$.  The $\grpu{n+1}$ generators are given in tensor
form by $(s^\dagger\times\tilde{s})^{(0)}$,
$(s^\dagger\times\tilde{b})^{(L)}$,
$(b^\dagger\times\tilde{s})^{(L)}$, and
$(b^\dagger\times\tilde{b})^{(\lambda)}$ (see
Appendix~\ref{secappboson} for detailed definitions).  If the
Hamiltonian is simply taken as the Casimir operator~(\ref{eqncasimir})
of either of the subalgebras, $\grpso{n+1}$ or $\grpu{n}$, a dynamical
symmetry is obtained.  The $\grpu{n}$ symmetry is geometrically
related to the $n$-dimensional harmonic oscillator, the $\grpso{n+1}$
symmetry to the $n$-dimensional rotator-vibrator (\textit{e.g.},
Ref.~\cite{iachello2006:liealg}).

The ground state QPT in the two-level boson models arises as the
Hamiltonian is varied linearly between the two dynamical
symmetries, for instance, by varying $\xi$ in the Hamiltonian
\begin{equation}
\label{eqnHMMxi}
\Hhat= \frac{(1-\xi)}{N} \Nhat_b - \frac{\xi}{N^2} (s^\dagger \tilde{b} + b^\dagger
\tilde{s})\cdot(s^\dagger \tilde{b} + b^\dagger \tilde{s}),
\end{equation}
where $\Nhat_b\<\equiv (-)^L b^\dagger \cdot \tilde{b}$ is the
$b$-boson occupancy, $\tilde{T}^{(\lambda)}_\mu\<\equiv(-)^{\lambda-\mu}T^{(\lambda)}_{-\mu}$,
and $U^{(\lambda)}\cdot V^{(\lambda)}\<\equiv(-)^L
(2L+1)^{1/2}(A\times B)^{(0)}$.  This Hamiltonian yields the $\grpu{n}$ symmetry
for $\xi\<=0$ and the $\grpso{n+1}$ symmetry for $\xi\<=1$.  The
Hamiltonian is invariant under the common $\grpso{n}$ algebra
in~(\ref{eqnchainuso}) and therefore conserves a $(2L+1)$-dimensional
angular momentum quantum number $v$.  As $\xi$ is increased from
$\xi\<=0$, the increasing strength of the interaction between $s$ and
$b$ levels changes the structure of the ground state from a pure
$s$-boson condensate to a condensate involving both types of bosons.
For asymptotically large values of the total particle number
$N\<\equiv N_s+N_b$, the change is abrupt.  A second-order ground
state QPT is well known to occur for $\xi_c\<=1/5$, with all the
properties enumerated in Sec.~\ref{secintro}.  [The coefficients
in~(\ref{eqnHMMxi}) are scaled by appropriate powers of $N$ to guarantee
that the location of the critical point is independent of $N$ in the
large $N$ limit.]  With more complex interactions in the Hamiltonian,
first-order QPTs, such as the physically important
$\grpu{5}$--$\grpsu{3}$ phase transition in the IBM, may also be
obtained~\cite{dieperink1980:ibm-classical,feng1981:ibm-phase,arias2006:two-level-holstein-primakoff}.
The conditions under which such first-order phase transitions occur in
an arbitrary $\grpu{n+1}$ model are outlined in
Ref.~\cite{cejnar2007:phase-cusp-un}.  However, only second-order
ground state QPTs will be considered here.

We observe, moreover, that the $\grpu{n+1}$ two-level boson models are
special cases of an even larger family of models, the two-level
pairing models with quasispin Hamiltonians.  Two-level pairing models
can be defined for systems of either bosons 
or fermions.  The two-level pairing models
undergo a second-order \textit{ground state}
QPT~\cite{broglia1968:pairing-transition}.  Therefore, it is natural
to consider the possibility that \textit{excited state} QPTs may occur
within the context of this broader family of models as well.

The quasispin pairing Hamiltonian is of the form
\begin{equation}
\label{eqnHquasi}
\Hhat=\sum_j\varepsilon_j \Bigl(\sum_m c_{jm}^\dagger \tilde{c}_{jm}
\Bigr) + \frac{1}{4}\sum_{j'j}G_{j'j}
\Bigl(\sum_{m'} c_{j'm'}^\dagger \tilde{c}_{j'm'}^\dagger \Bigr)
\Bigl(\sum_m \tilde{c}_{jm} c_{jm} \Bigr),
\end{equation}
where the summation indices $j$ and $j'$ run over the single-particle
levels, and $m$ and $m'$ run over their
substates.  The $c_{j}$ may represent either bosonic operators
$b_1^{(L_1)}$ and $b_2^{(L_2)}$ [Fig.~\ref{fig-levels}(b)] or fermionic
operators operators $a_1^{(j_1)}$ and $a_2^{(j_2)}$
[Fig.~\ref{fig-levels}(c)], as appropriate.  Although the
Hamiltonian~(\ref{eqnHquasi}) superficially appears quite different
from the $\grpu{n+1}$ two-level boson model
Hamiltonian~(\ref{eqnHMMxi}), the two are in fact
equivalent~\cite{arima1979:ibm-o6,pan1998:so6u5}.  The detailed
relationship between the models is established for arbitrary $n$ in
Appendix~\ref{secappboson}.

It is well known that the pairing Hamiltonian~(\ref{eqnHquasi}) can be
expressed in terms of the generators $\Shat_{j+}$, $\Shat_{j-}$, and
$\Shat_{jz}$ of a quasispin algebra~(\ref{eqnScomm}), as
\begin{equation}
\label{eqnHquasiSS}
\Hhat=\sum_j\varepsilon_j (2\Shat_{jz}\mp\Omega_j) + \sum_{j'j}G_{j'j}\Shat_{j'+}\Shat_{j-},
\end{equation}
where $\Omega_j\<\equiv (2j+1)/2$, and the upper and lower signs apply
in the bosonic and fermionic cases, respectively.  The algebra is
either an $\grpsu{1,1}$ algebra if the operators are
bosonic~\cite{ui1968:su11-quasispin-shell} or an $\grpsu{2}$ algebra
if the operators are fermionic~\cite{kerman1961:pairing-collective}.
However, the pairing models are also characterized by an overlaid
$\grpu{n_1+n_2}$ algebraic structure, described further in
Ref.~\cite{caprio:two-level-alg}, either
\begin{equation}
\label{eqnchainu12so12}
\grpu{n_1+n_2}
\supset
\left\lbrace
\begin{array}{c}
\grpso{n_1+n_2}\\
\grpu[1]{n_1}\otimes\grpu[2]{n_2}
\end{array}
\right\rbrace
\supset
\grpso[1]{n_1}\otimes\grpso[2]{n_2}
\supset
\grpso[12]{3}
\end{equation}
in the bosonic case (with $n_1\<=2L_1+1$ and $n_2\<=2L_2+1$) or
\begin{equation}
\label{eqnchainu12sp12}
\grpu{n_1+n_2}
\supset
\left\lbrace
\begin{array}{c}
\grpsp{n_1+n_2}\\
\grpu[1]{n_1}\otimes\grpu[2]{n_2}
\end{array}
\right\rbrace
\supset
\grpsp[1]{n_1}\otimes\grpsp[2]{n_2}
\supset
\grpsu[12]{2}
\end{equation}
in the fermionic case (with $n_1\<=2j_1+1$ and $n_2\<=2j_2+1$),
directly generalizing the $\grpu{n+1}$ algebraic structure~(\ref{eqnchainuso}) of
the $s$-$b$ boson models.  The $\grpu{n_1+n_2}$ generators are of
the form $(c_1^\dagger\times\tilde{c}_1)^{(\lambda)}$,
$(c_1^\dagger\times\tilde{c}_2)^{(\lambda)}$,
$(c_2^\dagger\times\tilde{c}_1)^{(\lambda)}$, and
$(c_2^\dagger\times\tilde{c}_2)^{(\lambda)}$.  The $\grpso[1]{n_1}$ and
$\grpso[2]{n_2}$ [or $\grpsp[1]{n_1}$ and $\grpsp[2]{n_2}$] algebras
provide conserved $n_1$-dimensional and $n_2$-dimensional angular
momentum quantum numbers ($v_1$ and $v_2$), which are equal to the
seniority quantum numbers defined in the quasispin formulation.

The ground state QPT in the general two-level pairing models is
between the $\grpso{n_1+n_2}$ or $\grpsp{n_1+n_2}$ dynamical symmetry
and the $\grpu[1]{n_1}\otimes\grpu[2]{n_2}$ dynamical symmetry.
To choose a transitional Hamiltonian for the general pairing models
consistent with the Hamiltonian already used for the $s$-$b$
boson models, we observe that the Hamiltonian~(\ref{eqnHMMxi}) may be
reexpressed (see Appendix~\ref{secappboson}) in pairing form as
\begin{equation}
\label{eqnHPPxi}
\Hhat= \frac{(1-\xi)}{N} \Nhat_2 + \frac{4\xi}{N^2} (-)^{L+1} (\Shat_{1+}\pm\Shat_{2+}) (\Shat_{1-}\pm\Shat_{2-}),
\end{equation}
to within an additive constant, where the full relation is
given explicitly in~(\ref{eqnHdiff}).  With this
form of Hamiltonian for the pairing models, the ground state QPT again
occurs at $\xic\<=1/5$.

Since QPTs occur in the limit of large particle number, an important
distinction arises between bosonic and fermionic models.  Arbitrarily
large particle number can be achieved in the \textit{bosonic} models,
even for fixed level degeneracies, simply by increasing the total
occupancy.  For a \textit{fermionic} model, however, the total
occupancy is limited by Pauli exclusion to the total degeneracy
[$(2j_1+1)+(2j_2+1)$].  Therefore, the limit of large particle number
can only be achieved if the number of available substates in each
level is simultaneously increased.  For two fermionic levels of equal
degeneracy ($j_1\<=j_2\<\equiv j$), half-filling is achieved for
$N=2j+1$.

It will be convenient to make extensive use of the $\grpu{3}$
two-dimensional vibron
model~\cite{iachello1996:vibron-2dim,perez-bernal2005:vibron-2dim-nonrigidity,perez-bernal:u3-vibron-phase}
for illustration in this article.  The $\grpu{3}$ vibron model is the
simplest two-level model which still retains a nontrivial angular
momentum or seniority quantum number (unlike the Lipkin model).
\begin{figure}
\begin{center}
\includegraphics*[width=0.6\hsize]{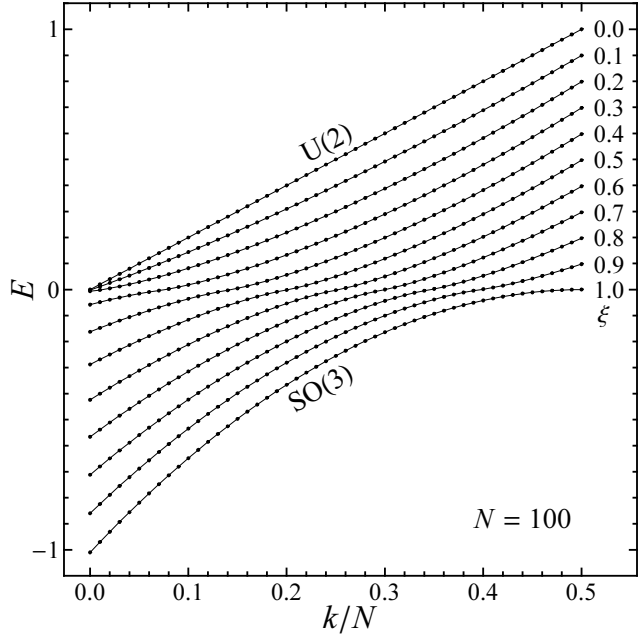}
\end{center}
\caption{Eigenvalue spectra for the $\grpu{3}$ vibron model
$l\<=0$ states ($N\<=100$), for several specific values of the Hamiltonian
parameter $\xi$.  Eigenvalues are plotted with respect to the scaled
excitation quantum number $k/N$.  
}
\label{fig-spectra}
\end{figure}

First, note that single-particle levels in bosonic pairing models
[Fig.~\ref{fig-levels}(a,b)] are only restricted to odd degeneracies
(\textit{i.e.}, $2L+1$ with $L$ integer) if a physical
three-dimensional angular momentum subalgebra $\grpso{3}$ is required
in~(\ref{eqnchainuso}) or~(\ref{eqnchainu12so12}).  The pairing
interaction only requires the definition of time-reversed pairs.  It
therefore suffices to have an ``$M$'' quantum number, with pairs $\pm
M$, without necessity for an ``$L$'' quantum number.
The pairing interaction can therefore be defined for an even number of
bosons, and the interaction within each level is described by
$\grpso{n}$ with $n$ even.\footnote{However, pairing is not
well-defined for the converse situation, a fermionic level of odd
degeneracy.  With an odd number of substates, one (``$m\<=0$'') must
necessarily be its own conjugate under time reversal.  Creation of a
time-reversed pair involving this substate is Pauli forbidden.  The
corresponding algebra, $\grpsp{n}$ with $n$ odd, is not defined.}
Bosonic levels of even degeneracy arise naturally in problems lacking
three-dimensional rotational invariance.

The $\grpu{3}$ vibron model may be obtained by considering the $\grpu{4}$
vibron model ($L\<=1$) and eliminating the substate $b^{(1)}_0$.  This
leaves a $\grpu{3}$ algebraic structure, with $\grpso{3}$ and
$\grpu{2}$ dynamical symmetries.  The geometrical coordinates
associated with the $\grpu{4}$ model describe three-dimensional dipole
motion (as in a linear dipole molecule).  However, eliminaton of
$b^{(1)}_0$ ``freezes out'' motion in the $z$ direction, so the
$\grpu{3}$ model instead describes two-dimensional motion in the $xy$
plane.  The $\grpu{2}$--$\grpso{3}$ transitional Hamiltonian, in
Casimir form,
is~\cite{iachello1996:vibron-2dim,perez-bernal2005:vibron-2dim-nonrigidity},
\begin{equation}
\label{eqnHvibron}
H
=
\frac{(1-\xi)}{N}\Nhat_b-\frac{\xi}{N^2}\bigl[\tfrac12(\Dhat_+\Dhat_-+\Dhat_-\Dhat_+)+\lhat^2\bigr],
\end{equation}
where $\Dhat_\pm\<\equiv\pm\sqrt{2}(b_{\pm1}^\dag s_0 -s_0^\dag
b_{\mp1})$.  This is the two-dimensional equivalent
of~(\ref{eqnHMMxi}), to within an additive constant.  The conserved
two-dimensional angular momentum is $\lhat\<= b_{+1}^\dag b_{+1} -
b_{-1}^\dag b_{-1}$.  The eigenvalue spectra of $l\<=0$ states, for
various values of $\xi$, are shown in Fig.~\ref{fig-spectra}.  The
spectra for the $\grpu{2}$ dynamical symmetry ($\xi\<=0$) and the
$\grpso{3}$ dynamical symmetry ($\xi\<=1$) have simple analytic
forms~\cite{iachello1996:vibron-2dim}.  Note also the spectrum for the
ground state QPT ($\xi\<=0.2$).

%%%%%%%%%%%%%%%%%%%%%%%%%%%%%%%%%%%%%%%%%%%%%%%%%%%%%%%%%%%%%%%%
%%%%%%%%%%%%%%%%%%%%%%%%%%%%%%%%%%%%%%%%%%%%%%%%%%%%%%%%%%%%%%%%

\section{Semiclassical dynamics}
\label{secclass}

\subsection{Coordinate Hamiltonian}
\label{secclasscoord}

Each of the many-body models considered in Sec.~\ref{secmodels} has an
associated classical Hamiltonian, defined with respect to classical
coordinates and momenta, which is obtained through the use of coherent
states~\cite{gilmore1974:lie-groups,feng1981:ibm-phase,zhang1990:coherent}.
The basic properties of the excited state quantum phase transition
follow from the semiclassical analysis of a double-well potential with
a parabolic barrier [Fig.~\ref{fig-class-action}(c)] or, in higher
dimensions, a sombrero potential (also known as the ``champagne
bottle'' potential~\cite{child1998:champagne-monodromy}).  The
semiclassical dynamics for these potentials has been studied in
depth~\cite{ford1959:parabolic-barrier,child1998:champagne-monodromy,cary1987:separatrix-eigenvalues,cary1992:separatrix-eigenfunctions,cary1993:separatrix-dynamics},
and the connection with ESQPT phenomena in the Lipkin model and
higher-dimensional $s$-$b$ boson models has been made in
Refs.~\cite{heiss2002:qpt-instability,leyvraz2005:lipkin-scaling,macek2006:ibm-o6-u5-part2-classical-trajectories,cejnar2006:excited-qpt}.
In particular, at the energy of the top of the barrier, the classical
action undergoes a logarithmic singularity, which leads
semiclassically to the prediction of an infinite level density.  Here
we do not attempt a comprehensive recapitulation of the existing
analysis but rather briefly summarize the essential points and derive some results specifically relevant to the
observables of interest in phase transitional phenomena.

For the quasispin models of Sec.~\ref{secmodels}, the two superposed
algebraic structures (quasispin and unitary) give rise to two
alternative sets of coherent states and therefore to two realizations
of the classical dynamics.  The $\grpsu{1,1}$ or $\grpsu{2}$ quasispin
algebra yields a one-dimensional dynamics (the phase space is a Bloch
sphere or hyperboloid~\cite[Ch.~6]{gilmore1974:lie-groups}) which is
common to all the quasispin models.  The dynamics arising from the
quasispin algebra therefore highlights aspects universal to these
models, yielding the basic double-well potential
[Fig.~\ref{fig-class-action}(c)] and therefore indicating that all
should exhibit an ESQPT at the energy of the top of the barrier.  In
contrast, the coherent states obtained from the unitary
$\grpu{n_1+n_2}$ algebra yield a much richer classical dynamics, in
$n_1n_2$ dimensions, associated with the coset space
$\grpu{n_1+n_2}/[\grpu{n_1}\otimes\grpu{n_2}]$~\cite[Ch.~9]{gilmore1974:lie-groups}.
This more complete dynamics, so far only fully investigated for the
$s$-$b$
models~\cite{ginocchio1980:ibm-coherent-bohr,hatch1982:ibm-classical},
yields a much more detailed description of the system.  The dynamics
obtained from the quasispin algebra is essentially a one-dimensional
projection or ``shadow'' of the full dynamics arising from the unitary
algebra, as described by Feng, Gilmore, and
Deans~\cite{feng1981:ibm-phase} for the IBM.  In particular, the
presence of angular degrees of freedom and conserved angular momentum
quantum numbers have significant consequences for the
ESQPT~\cite{macek2006:ibm-o6-u5-part2-classical-trajectories,cejnar2006:excited-qpt}.

First, let us summarize the classical Hamiltonian obtained from the
$\grpu{n+1}$ coherent states for the $s$-$b$ model.  The classical
Hamiltonian acts on $n$ coordinates and their conjugate momenta.
However, for the $\grpso{n}$-invariant interaction
in~(\ref{eqnHMMxi}), the Hamiltonian is invariant under rotations in
the $n$-dimensional space and can therefore be expressed solely in
terms of a radial coordinate $r$, its conjugate momentum $p_r$, and
a conserved angular kinetic energy $T_\vartheta(v)$,
as~\cite{vanroosmalen1982:diss,hatch1982:ibm-classical,cejnar2006:excited-qpt}
\begin{equation}
\label{eqnHclassr}
\Hhat=\frac{1-\xi}{2N^2}[p_r^2+r^{-2}T_\vartheta(v)]+\frac{\xi}{N^2}[r^2p_r^2+T_\vartheta(v)]
+\frac{1-5\xi}{2}r^2+\xi r^4,
\end{equation}
where $T_\vartheta(v)$ has eigenvalue $v(v+n-2)$ and the coordinate
$r$ is defined only on the domain $0\<\leq
r\<\leq\sqrt{2}$.\footnote{In obtaining~(\ref{eqnHclassr}) from
Ref.~\cite{hatch1982:ibm-classical}, a scaling transformation
$r\<\rightarrow N^{1/2} r$ has been made, and a constant term of
order $1/N$ has been suppressed.}  The eigenvalue problem
for~(\ref{eqnHclassr}) therefore has the form of a \textit{radial}
Schr\"odinger equation with a quadratic-quartic potential, except for
the appearance of the position-dependent kinetic energy term
proportional to $r^2p_r^2$.  For the one-dimensional case,
\textit{i.e.}, the Lipkin model, the centrifugal term is not present,
and the coordinate and momentum are more aptly denoted by $x$ and $p$,
so
\begin{equation}
\label{eqnHclassx}
\Hhat=\frac{1-\xi}{2N^2}p^2+\frac{\xi}{N^2}x^2p^2
+\frac{1-5\xi}{2}x^2+\xi x^4,
\end{equation}
where here both negative and positive values of the coordinate $x$ are
allowed ($-\sqrt{2}\<\leq x\<\leq+\sqrt{2}$).

The role of $\hbarfracinline$ in the usual Schr\"odinger equation is
taken on by the coefficient of $p_r^2$ or $p^2$ in~(\ref{eqnHclassr})
or~(\ref{eqnHclassx}).  We therefore make the identification
$\hbar\<\rightarrow N^{-1}$, with the coordinate-dependent mass
$m(x)\<=(1-\xi+2\xi x^2)^{-1}$.

The forms
assumed by the quadratic-quartic potential in~(\ref{eqnHclassr})
or~(\ref{eqnHclassx}),
\begin{math}
\label{eqnVclass}
V(x)\<=(1-5\xi)x^2/2+\xi x^4,
\end{math}
are summarized for convenience in Fig.~\ref{fig-class-action}(a--c).
For the radial problem, of course, only the positive abscissa is
relevant.  For $\xi\<<1/5$, the potential has a single minimum, at
$x\<=0$, which is locally quadratic.  For $\xi\<=1/5$, the critical
value for the ground state QPT, the potential is pure quartic.  For
$\xi\<>1/5$, the familiar double-well potential is obtained (or the
sombrero potential for $n\<>1$).  For the
Hamiltonian~(\ref{eqnHclassr}) or~(\ref{eqnHclassx}), the zero in
energy is such that the top of the barrier is always at $E\<=0$,
independent of $\xi$.  

\subsection{Singular properties of the action}
\label{secclasssing}

The main semiclassical features of levels at energies near the top of
the barrier are obtained by noting that for $E\<=0$ the classical
velocity $v(E,x)\<=[2[E-V(x)]/m(x)]^{1/2}$ locally vanishes at the top
of the barrier ($x\<=0$).  While indeed the classical velocity also
vanishes at the ordinary linear turning points of a potential well,
the vanishing slope at the top of the barrier presents a qualitatively
broader ``flat'' region over which the classical velocity is small.
Thus, the semiclassical motion has a long ``dwell time'' in the
vicinity of $x\<=0$.  This leads to two essential results, namely
(1)~an infinite period $\tau\<=\oint v(E,x)^{-1}\,dx$ for classical
motion across the top of the barrier and (2)~strong localization of
the semiclassical probability density $P(x)\<\propto v(E,x)^{-1}$ at
the top of the barrier~\cite{leyvraz2005:lipkin-scaling}.

\begin{figure}
\begin{center}
\includegraphics*[width=0.8\hsize]{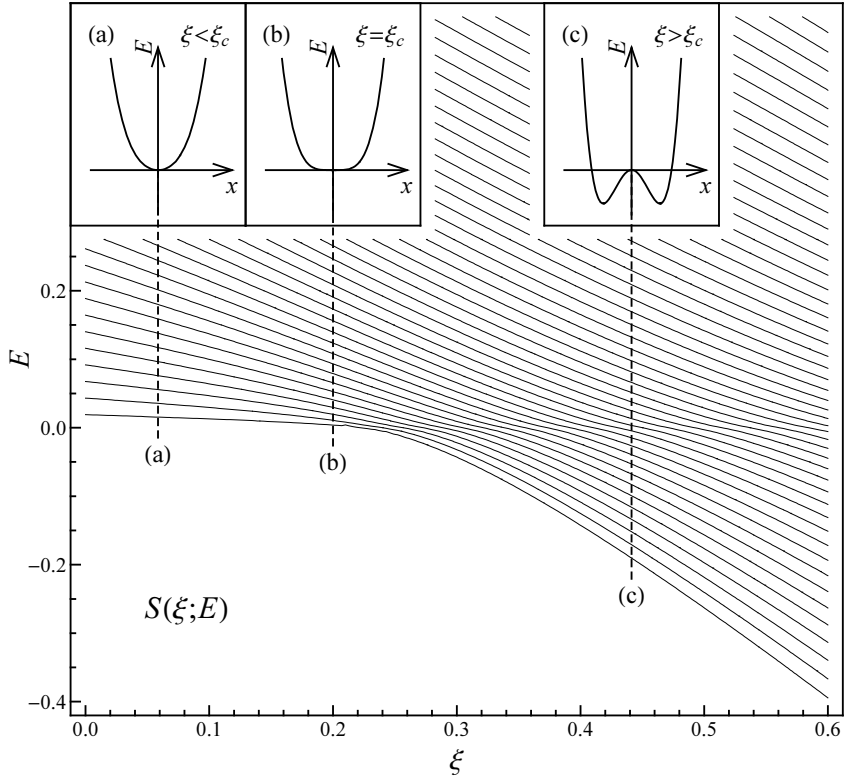}
\end{center}
\caption{
Contour plot showing the global structure of the classical action
$S(\xi;E)$ for the geometric Hamiltonian~(\ref{eqnHclassr})
or~(\ref{eqnHclassx}), through the different regimes determined by the
shape of the quadratic-quartic potential energy
function~(\ref{eqnVclass}), which is shown for (a)~$\xi\<<\xic$,
(b)~$\xi\<=\xic$, and (c)~$\xi\<>\xic$.  The individual contours are
related semiclassically to the evolution of the level eigenvalues
$E_k(\xi)$.  }
\label{fig-class-action}
\end{figure}

The first-order semiclassical analysis provides a simple guidemap to
the properties of the spectrum as a whole and also provides an
explanation for the singularity in level density as the top of the
barrier is approached.  We consider the one-dimensional
problem~(\ref{eqnHclassx}), but the results apply equally to the
radial problem~(\ref{eqnHclassr}) with $v\<=0$.  
For the Hamiltonian~(\ref{eqnHclassx}), the usual first-order WKB 
quantization condition~\cite{messiah1999:qm} becomes
\begin{equation}
\label{eqnWKBquant}
S(\xi;E)=(k+\tfrac12)2\pi N^{-1},
\end{equation}
with $k\<=0$, $1$, $\ldots$, where the action $S\<\equiv\oint p\,dx$ over a full classical
period of motion is given by the integral
\begin{equation}
\label{eqnS}
S(\xi;E) = 2 \int_{x_1(E)}^{x_2(E)} dx \, \bigl[ 2 m(x) [E-V(x)]
\bigr]^{1/2}
\end{equation}
between classical turning points $x_1(E)$ and $x_2(E)$.\footnote{Some
bookkeeping issues naturally must be taken into account in the
one-dimensional double-well problem [Fig.~\ref{fig-class-action}(c)].
For $E\<<0$, \textit{i.e.}, below the barrier, the two wells are
classically isolated.  Applying the quantization condition with $S$
evaluated over one of the wells in isolation is equivalent to counting
only states of one parity (symmetric or antisymmetric).  For $E\<>0$,
applying the quantization condition with $S$ evaluated over the full
well counts states of both parity.  Questions as to the proper
transition between the regimes $E\<<0$ and $E\<>0$ are somewhat
artificial, since the validity conditions for~(\ref{eqnWKBquant})
break down at $E\<\approx0$.}  The action depends upon $\xi$ variously
through $m(x)$, $V(x)$, and the turning points.

\begin{figure}
\begin{center}
\includegraphics*[width=0.6\hsize]{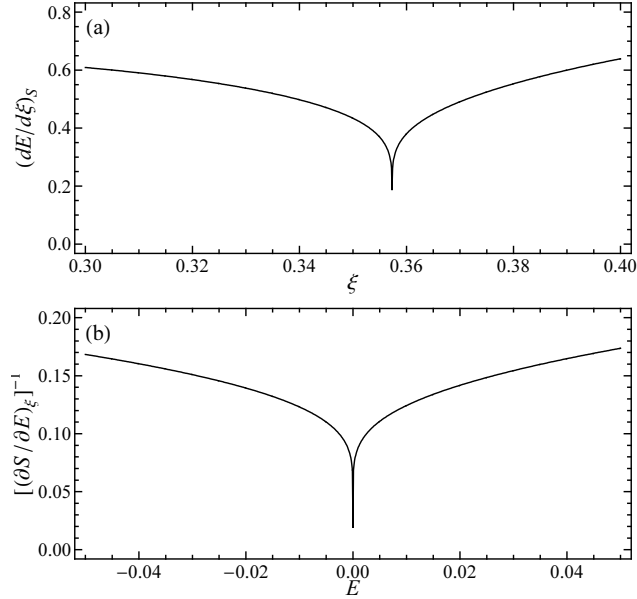}
\end{center}
\caption{Singularities in derivatives of the classical action~(\ref{eqnS}) for the
geometric Hamiltonian~(\ref{eqnHclassr}) or~(\ref{eqnHclassx}).
(a)~The derivative $dE/d\xi$ along a countour of $S(\xi;E)$
(Fig.~\ref{fig-class-action}), related semiclassically to the
adiabatic evolution of the level energy. (b)~The inverse of the
partial derivative $(\partial S/ \partial E)_\xi$, proportional to
the semiclassical estimate for the gap.
}
\label{fig-class-deriv}
\end{figure}

The quantization condition~(\ref{eqnWKBquant}) implicitly gives the
adiabatic evolution of the energy $E_k(\xi)$ of a given level with
respect to the parameter $\xi$.  Since~(\ref{eqnWKBquant}) enforces
that $S(\xi;E)$ be constant if $k$ is held fixed, the curve describing
$E_k(\xi)$ is simply a contour of $S(\xi;E)$ in the $\xi$-$E$ plane.
These contours, calculated numerically for the
Hamiltonian~(\ref{eqnHclassx}) [or~(\ref{eqnHclassr}) with $v\<=0$]
are plotted in Fig.~\ref{fig-class-action}.  A compression of
energy levels at $E\<=0$ is visible qualitatively even here.  [In
Fig.~\ref{fig-evoln-pairing}, the $E_k(\xi)$ are plotted as
\textit{excitation} energies and therefore cannot be compared
directly with Fig.~\ref{fig-class-action}.  More appropriate plots
for comparison may be found in the following section, \textit{e.g.},
Fig.~\ref{fig-evoln-l}(a).]  The derivative $dE_k/d\xi$ along a
single contour of $S(\xi;E)$ is plotted in Fig.~\ref{fig-class-deriv}(a).
Note that $dE_k/d\xi$ undergoes a singularity in which
$dE_k/d\xi\<\rightarrow 0$ but
$d^2E_k/d\xi^2\<\rightarrow \pm \infty$, at a critical value $\xi\<=\xicex$.

In semiclassical analysis, the gap or level density is directly
related to the classical period.  From the quantization
condition~(\ref{eqnWKBquant}), it follows that the semiclassical
estimate of the gap between adjacent levels ($\Delta\<=dE_k/dk$) is
$\Delta(E)\<=2\pi N^{-1} (\partial S /\partial E)^{-1}$.  By
differentiation of~(\ref{eqnS}), the gap is simply $\Delta(E)\<=2\pi
N^{-1} \tau^{-1}$.  As already noted for the
ESQPT~\cite{leyvraz2005:lipkin-scaling}, the period $\tau$ becomes
infinite at $E\<=0$ and, equivalently, the gap $\Delta(E)$ vanishes.
An explicit calculation of $(\partial S /\partial E)^{-1}$ as a
function of $E$ for the classical Hamiltonian~(\ref{eqnHclassx}) is
shown in Fig.~\ref{fig-class-deriv}(b).  Note that $(\partial S
/\partial E)^{-1}$ undergoes a singularity in which $(\partial S
/\partial E)^{-1}\<\rightarrow 0$ but $(\partial^2 S /\partial
E^2)^{-1}\<\rightarrow \pm \infty$, at the critical energy $\Ec\<=0$.

For nonzero angular momentum $v$, the origin ($r\<=0$) is classically
forbidden due to the centrifugal term in~(\ref{eqnHclassr}), which
causes the wave function probability near the origin to be suppressed.
This mitigates the effects just described, by masking the top of the
barrier and precluding the long semiclassical dwell time at the
origin~\cite{cejnar2006:excited-qpt}.  The dependence of the
Hamiltonian~(\ref{eqnHclassr}) on $v$ is through the coefficient of
the centrifugal term, which is proportional to
$T_\vartheta(v)/N^2\<\approx (v/N)^2$.  Therefore, the phenomena
associated with the ESQPT can be expected to be suppressed for
sufficiently large $v$ at any \textit{given} value of $N$.  On the
other hand, the angular momentum effects at any \textit{given} value
of $v$ are negligible for sufficiently large $N$.  That is, the
signatures of the ESQPT persist for small $v$ ($v/N\<<<1$) and only
disappear for $v/N$ of order unity (as illustrated quantitatively in
Sec.~\ref{secobseigen}).

\subsection{Asymptotic spectrum}
\label{secclassasymp}

Let us now consider more precisely the form of the singularity in the
spectrum in the immediate neighborhood of the ESQPT.  As the
wave function becomes increasingly well-localized near the top of the
barrier for $E\<\rightarrow0$, it should become an increasingly good
approximation to treat the barrier as a pure inverted oscillator
potential, $V(x)\<=-Ax^2$.  The position-dependent kinetic energy term
($\rtrim\propto x^2 p^2$) also becomes irrelevant.

In the action integral~(\ref{eqnS}), the classical turning point at
the barrier is $x_1(E)\<=(E/A)^{1/2}$ for $E\<<0$, or for $E\<>0$
integration simply extends to the origin.  The distant turning point
$x_2(E)$ is a slowly varying function of $E$ which does not contribute
to the singularity, so we may take it to be a constant.  (In any case,
for the actual potential, the approximation of a pure parabolic
barrier breaks down well before the distant turning point is reached.)
The action integral for $E\<>0$ is therefore
\begin{equation}
\begin{aligned}
\hbar^{-1} S(E)&=\frac{2}{\hbar} 
\int_{0}^{x_2} dx \, [2m(E+Ax^2)]^{1/2}
\\
&=\frac{4}{\hbar\omega}  E \int_0^{(A/E)^{1/2} x_2} du \,
(1+u^2)^{1/2},
\end{aligned}
\end{equation}
where, for the inverted oscillator Hamiltonian
\begin{math}
\Hhat\<=[\hbarfracinline] p^2 - A x^2,
\end{math}
we have defined 
\begin{math}
\hbar\omega \<= 2 [\hbarfracinline]^{1/2} A^{1/2}
\end{math}
by analogy with the conventional harmonic oscillator.

Expanding this action integral~\cite[(2.271.3)]{gradshteyn1994:table} for
$E\<\approx 0$ yields
\begin{equation}
\label{eqnSexpansion}
\hbar^{-1}S(E)=  \frac1{\hbar\omega} \bigl( -  E \log E +
\alpha_0 + \alpha E  + \cdots \bigr),
\end{equation}
where $\alpha_0$ and $\alpha$ are constants, \textit{i.e.}, depend
only on the potential parameters $A$ and $x_2$.  An essentially
identical result is obtained for
$E\<<0$, with the replacement
$E\<\rightarrow\abs{E}$~\cite[(1.646.2)]{gradshteyn1994:table}.  The singular behavior for energies near the
top of the barrier therefore arises from the $E
\log E$ term.\footnote{Since the Schr\"odinger equation for a
pure parabolic barrier is exactly solvable in terms of parabolic
cylinder functions~\cite{abramowitz1965}, the $\abs{E}\log\abs{E}$
dependence can also be obtained by explicitly matching this solution
for the wave function in the vicinity of the barrier to asymptotic WKB
wave functions away from the
barrier~\cite{thylwe2006:phase-integral-barrier}.}
The quantization condition~(\ref{eqnWKBquant}) takes on the form
\begin{equation}
\label{eqnElogEquant}
- E\log E + \alpha E + \cdots = 2\pi\hbar\omega(k-\kc),
\end{equation}
where $E\<=0$ is obtained for $k\<=\kc$.  If the energy dependence
in~(\ref{eqnElogEquant}) is truncated at the terms shown,
\textit{i.e.}, linear order in $E$, this quantization condition can be solved
for $E(k)$ in terms of the Lambert $W$ function, by~(\ref{eqnWylogy}),
yielding
\begin{math}
E(k)\<=-2\pi\hbar\omega(k-\kc)/W[-e^{-\alpha}2\pi\hbar\omega(k-\kc)]
\end{math}.
The relevant properties of the $W$ function are summarized in
Appendix~\ref{secapplambert}.

For the Hamiltonian~(\ref{eqnHclassx}), the top of the barrier is
described by an oscillator constant which may be read off from the
coefficients of the $p^2$ and $x^2$, giving
\begin{equation}
\label{eqnhbaromega}
\hbar\omega=\frac{\Xi(\xi)^{1/2}}{N}, 
\end{equation}
where
\begin{equation}
\label{eqnXi}
\Xi(\xi)\equiv(1-\xi)(1-5\xi).
\end{equation}
The oscillator constant thus
depends upon both $\xi$ and $N$.  The same function $\Xi(\xi)$,
interestingly, also enters into the \textit{ground state} QPT scaling
properties, as obtained by the continuous unitary transform method in
Ref.~\cite{dusuel2005:scaling-ibm}.  The semiclassical estimate for
the eigenvalue spectrum in the vicinity of the ESQPT is therefore
\begin{equation}
\label{eqnEW}
E(N,\xi,k)= - \frac{2\pi\Xi(\xi)^{1/2}(k-\kc)/N}{W[-e^{-\alpha}2\pi\Xi(\xi)^{1/2}(k-\kc)/N]},
\end{equation}
where $\alpha$ will contain a dependence on 
$\xi$ as well.  Differentiation with respect to $k$, making use
of~(\ref{eqnWderiv}), yields a semiclassical estimate
\begin{equation}
\label{eqnDeltaW}
\Delta(N,\xi,k)= - \frac{2\pi\Xi(\xi)^{1/2}/N}{W[-e^{-\alpha}2\pi\Xi(\xi)^{1/2}(k-\kc)/N]+1}
\end{equation}
for the energy gap between adjacent excited states.

Since the excitation quantum number and particle number enter into the
quantization condition~(\ref{eqnWKBquant}) together in the combination
$k/N$, the spectrum and finite-size scaling properties are
inextricably linked at the semiclassical level.\footnote{For the
\textit{ground state} QPT, the semiclassical potential is quartic
[Fig.~\ref{fig-class-action}(b)].  A simple application of the WKB
formula gives a dependence $E(k/N)\<\sim(k/N)^{4/3}$, which
simultanously yields both the spectrum $E_k\<\sim k^{4/3}$
[Fig.~\ref{fig-spectra} ($\xi\<=0.2$)] and the scaling $\Delta\<\sim
N^{-4/3}$ (Sec.~\ref{secobsscaling}).}  The expression~(\ref{eqnDeltaW}),
considered as a function of $N$ at fixed $k$, provides an estimate for
the scaling of the gap at the $(k-\kc)$-th eigenvalue above or below
$E\<=0$.  The large-$N$ behavior follows from the known asymptotic
form~(\ref{eqnWasymp}) of the $W$ function as a sum of logarithms
for $x\<\rightarrow 0^-$ (see Fig.~\ref{fig-lambertw}).  The values of
$x$ relevant to~(\ref{eqnDeltaW}) in the vicinity of the ESQPT are of the order
$x\<\sim-N^{-1}$.  The asymptotic form~(\ref{eqnWasymp}) provides a good
approximation to $W(x)$ for reasonable $N$, \textit{e.g.}, accurate to
$1\%$ by $N\<\sim10^5$.

For very large $N$, the scaling behavior is in principle even simpler.
The $\log (-x)$ term in~(\ref{eqnWasymp}) outgrows the $\log
[-\log(-x)]$ term as $x\<\rightarrow 0^-$. 
With this logarithmic approximation, an extreme asymptotic estimate
\begin{equation}
\label{eqnDeltalog}
N \Delta \sim - \frac{2\pi\Xi(\xi)^{1/2}}{\log (k-\kc) - \log N + \log
[2\pi\Xi(\xi)^{1/2}] - \alpha +1} \sim \frac{2\pi\Xi(\xi)^{1/2}}{\log N}
\end{equation}
is obtained, recovering the logarithmic scaling noted by Leyvraz and
Heiss~\cite{leyvraz2005:lipkin-scaling}.
However, even for $N\<\sim10^{10}$, the approximation $W(x)\<\sim
\log (-x)$ yields an error of $\rtrim>10\%$ and therefore is of
limited \textit{quantitative} value for systems of typical ``mesoscopic'' size.

Note that the quantization condition as given in~(\ref{eqnWKBquant})
is derived under the assumption that the classical turning points are
\textit{well separated} (by several de~Broglie wavelengths) and that
the potential is locally \textit{linear} at these turning
points~\cite{messiah1999:qm}.  This suffices for the analysis of levels
which are not close in energy to the top of the barrier.  However, for
$E\<\approx0$, the barrier presents a
\textit{quadratic} classical turning point.  (Equivalently, the linear
turning points on either side of the barrier approach each other,
violating the assumption of sufficient separation.)  For accurate quantitative
analysis of the levels immediately surrounding $E\<=0$, the more
general phase-integral method must be
applied~\cite{froeman2002:phase-integral}.  For a smooth, symmetric
double-well potential, the phase-integral method yields an approximate
quantization condition~\cite[(3.47.1)]{froeman2002:phase-integral}
\begin{equation}
\label{eqnphasequant}
\hbar^{-1} S(E)=2\pi(k+\tfrac12)-\tilde{\phi}+2\beta_0'' \pm \arctan \exp(-K),
\end{equation}
with $k$ an integer, where the various phases appearing on the right
hand side are defined in Ref.~\cite{froeman2002:phase-integral}.  The
full derivation involves the evaluation of contour integrals on the
complex extension of the coordinate axis and the consideration of
complex-valued turning points for energies just above the
barrier~\cite{froeman2002:phase-integral}.  Quantitative solution of
the problem is considered in detail in
Refs.~\cite{tennyson1986:separatrix-adiabatic,cary1986:separatrix-adiabatic,cary1987:separatrix-eigenvalues,cary1993:separatrix-dynamics}.
The effects of these corrections~(\ref{eqnphasequant}) relative
to~(\ref{eqnWKBquant}) are explored in
Ref.~\cite{child1998:champagne-monodromy}.  The corrections are
essential to the treatment of the first few eigenvalues above or below
the barrier.  However, here we are instead interested in extracting
the basic nature of the singularity from the dependence of $S(E)$ on
$E$ in the vicinity of $E\<=0$, for which the simple quantization
condition~(\ref{eqnWKBquant}) suffices.

%%%%%%%%%%%%%%%%%%%%%%%%%%%%%%%%%%%%%%%%%%%%%%%%%%%%%%%%%%%%%%%%
%%%%%%%%%%%%%%%%%%%%%%%%%%%%%%%%%%%%%%%%%%%%%%%%%%%%%%%%%%%%%%%%

\section{Quantum properties}
\label{secobs}

\subsection{Eigenvalue spectrum}
\label{secobseigen}

In a ground state QPT, the singular behavior of the system is
simultaneously reflected in the eigenvalue spectrum (ground state
energy and gap) and in the order parameters.  From the preceding
semiclassical analysis (Sec.~\ref{secclass}), it is to be expected
that a similar variety of interconnected phenomena occur at the ESQPT,
and this is indeed borne out by the quantum calculations.  Of course,
the analogy between ground state QPT and ESQPT is far from exact, so
let us now examine the results for spectra and order parameters
obtained numerically from the full quantum calculation, to elucidate
both the analogy with the ground state QPT and the applicability of
the semiclassical results of Sec.~\ref{secclass}.

While the ground state QPT may only be traversed by varying a
Hamiltonian parameter, the locus of the ESQPT is a curve in the
two-parameter space defined by the Hamiltonian parameter $\xi$ and the
excitation energy (as along the dense band in
Fig.~\ref{fig-evoln-pairing}).  Therefore, the ESQPT may be crossed
either ``horizontally'', by varying $\xi$, or ``vertically'', by
varying the excitation quantum number $k$ (or, equivalently, the
energy $E$) of the level being examined.

The energy spectrum consists of the set of eigenvalues
$E_{N,\xi,k,\Lambda}$, which contain dependences on several
quantities: the system size $N$, the Hamiltonian parameter $\xi$, the
excitation quantum number $k$, and other conserved quantum numbers
$\Lambda$ (angular momenta or seniorities in the present models).  For
large $N$, however, $k/N$ and $\Lambda/N$ become essentially
continuous variables.  In the preceding section, it was seen that
semiclassically the energy depends upon the quantum numbers only
through these combinations $k/N$ and $\Lambda/N$.  We are therefore
largely interested in the properties of the spectrum given by the
function $E(\xi,k/N,\Lambda/N)$ of three quasi-continuous
variables~\cite{vanroosmalen1982:diss}.  The dependence of the
spectrum on \textit{interaction}, \textit{excitation quantum number}, and
\textit{angular momentum} is contained in the dependence of 
$E(\xi,k/N,\Lambda/N)$ on its three arguments.\footnote{In the 
Hamiltonians~(\ref{eqnHMMxi}) and~(\ref{eqnHPPxi}),
the coefficients of the one-body operators are scaled by
$N$ and the coefficients of the two-body operators are scaled by
$N^2$.  Often a Hamiltonian normalization
differing by an overall factor of $N$ is instead used, \textit{e.g.},
\begin{math}
\Hhat= (1-\xi) \Nhat_b - (\xi/N) (s^\dagger \tilde{b} + b^\dagger
\tilde{s})\cdot(s^\dagger \tilde{b} + b^\dagger \tilde{s})
\end{math}
for the $s$-$b$ model. 
For the
normalization~(\ref{eqnHMMxi}) or~(\ref{eqnHPPxi}),
$E_{N,\xi,k,\Lambda}$ does indeed approach a limiting value
$E(\xi,k/N,\Lambda/N)$ as $N\<\rightarrow\infty$,
by~(\ref{eqnWKBquant}).  However, for the alternate normalization it
is actually $N^{-1} E_{N,\xi,k,\Lambda}$ which approaches a limiting
value as $N\<\rightarrow\infty$.}  Furthermore, note that the
dependence on the argument $k/N$ implicitly contains information not
only on the excitation spectrum [when the function is considered as
$E(k)$ at fixed $N$] but also on the
\textit{finite-size scaling behavior} [when the function is considered
as $E(1/N)$ at fixed $k$].  The properties of $E(\xi,k/N,\Lambda/N)$
in the vicinity of the ground state, that is, for $k/N\<\ll1$, have
been studied in detail, at least for the $s$-$b$ models.  Here,
instead, we are considering the regime $k/N\<\sim 1$.

\begin{figure}
\begin{center}
\includegraphics*[width=0.8\hsize]{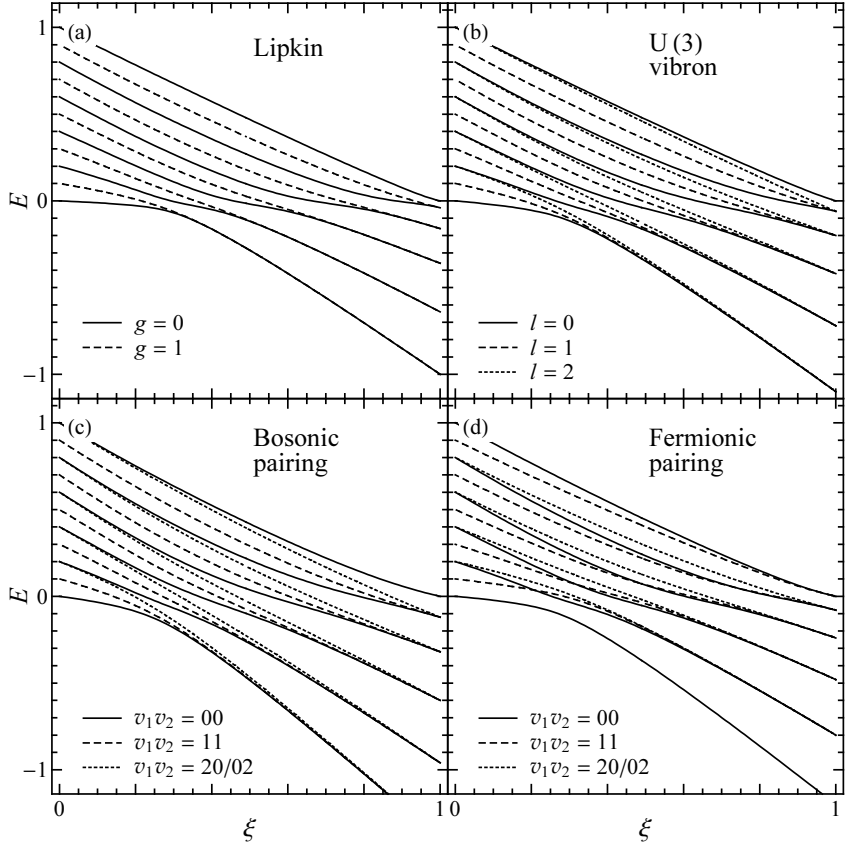}
\end{center}
\caption{
Eigenvalues for (a)~the Lipkin model (Schwinger realization), (b)~the
$\grpu{3}$ vibron model, (c)~the bosonic pairing model
($L_1\<=L_2\<=1$), and (d)~the fermionic pairing model
($j_1\<=j_2\<=9/2$), as functions of the coupling parameter $\xi$, all
for total particle number $N\<=10$.  For the Lipkin model, both
even-parity (solid curves) and odd-parity (dashed curves) levels are shown.  For the
other models, only the lowest angular momenta or seniorities are
shown.  A diagonal contribution $\xi(N+2L_1+2L_2)/N$ has been subtracted from the
Hamiltonian~(\ref{eqnHPPxi}) for the bosonic pairing model~\cite{caprio:two-level-alg}.}
\label{fig-evoln-models}
\end{figure}

First, let us establish the common ground between the various models
under consideration (Sec.~\ref{secmodels}), by a simple comparison of
the energy spectra.  Calculations are shown in
Fig.~\ref{fig-evoln-models} for the Lipkin model
[Fig.~\ref{fig-evoln-models}(a)], the $\grpu{3}$ vibron model
[Fig.~\ref{fig-evoln-models}(b)], a bosonic pairing model with equal
degeneracies for both levels ($L_1\<=L_2\<=1$)
[Fig.~\ref{fig-evoln-models}(c)], and a fermionic pairing model with
equal degeneracies ($j_1\<=j_2\<=9/2$)
[Fig.~\ref{fig-evoln-models}(d)].  The calculations are all for a
fixed, modest particle number ($N\<=10$), so that individual
eigenvalues are clearly distinguishable.  In the comparison, we must
distinguish the invariant subspaces of states for each model.  Each
eigenstate of the Lipkin model contains only even-$N_b$ or odd-$N_b$
components and is thus characterized by a grading quantum number
$g$ with values $0$ and $1$ ($g\<\cong N_b\mod{2}$) or,
equivalently, the parity $\pi\<=(-)^g$.  The vibron model states are
characterized by the angular momentum $l\<=0,\pm1,\ldots,\pm N$.  The
bosonic and fermionic pairing model states are characterized by
seniority quantum numbers for each single-particle
level, namely, $v_1$ and $v_2$.
\begin{figure}
\begin{center}
\includegraphics*[width=0.8\hsize]{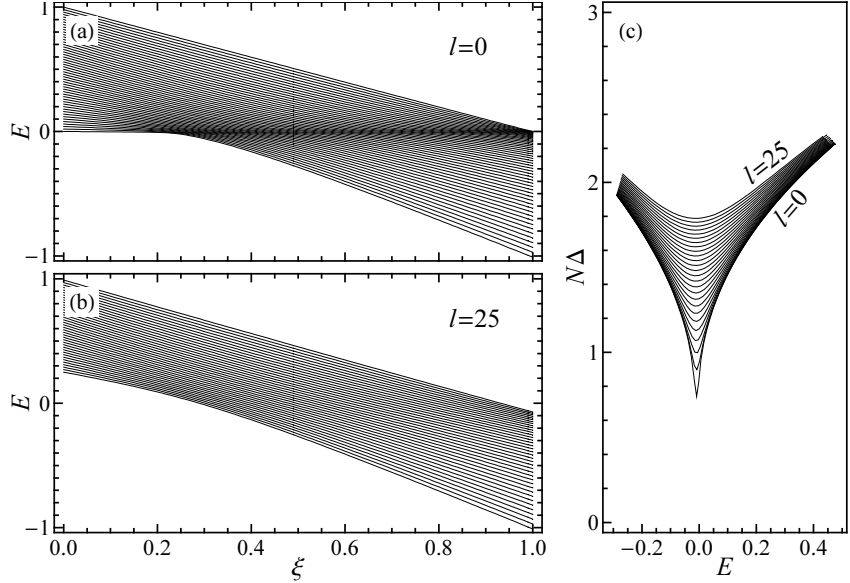}
\end{center}
\caption{
Angular momentum dependence of spectral properties for the
$\grpu{3}$ vibron model ($N\<=100$).  (a,b)~Evolution of eigenvalues
with $\xi$ for $l\<=0$ and $l\<=25$, \textit{i.e.}, $l/N\<=1/4$.  (c)~Dependence of
the gap on excitation energy, as in Fig.~\ref{fig-evoln-ex-combo}(b),
for various $l$ ($0\<\leq l \<\leq 25$).
}
\label{fig-evoln-l}
\end{figure}

Note the essentially identical evolution, with respect to $\xi$, of
the even-parity ($g\<=0$) states of the Lipkin model, the zero angular
momentum ($l\<=0$) states of the vibron model, and the zero seniority
[$(v_1v_2)\<=(00)$] states of both the bosonic and fermionic pairing
models (solid curves in Fig.~\ref{fig-evoln-models}).  The ground
state energy is near constant, with $E_0\<\approx0$, for $\xi\<<\xic$
and decreases to $\rtrim\lesssim-1$ for $\xi\<=1$.  The highest
eigenvalue decreases approximately linearly with $\xi$, from $1$ to
$0$.  Various qualitative features associated with the ESQPT occur at
$E\<\approx0$ for $\xi\<>\xic$ for these models.  Note especially the
inflection points for these levels (solid curves) as well as the
change in the pattern of degeneracies between different seniorities
(or parities or angular momenta) at $E\<\approx0$.

The major differences among the models lie in the degeneracy patterns
at nonzero seniority, which depend upon the specific algebraic
properties of the individual models~\cite{caprio:two-level-alg}.  At
present, we will limit consideration of angular momentum effects to
the $s$-$b$ models, since for these only one angular momentum quantum
number is involved, and $l$ in the $\grpu{3}$ vibron model
(Sec.~\ref{secmodels}) serves as a natural example for illustration.

The semiclassical analysis of Sec.~\ref{secclass} provided a simple
set of predictions (Fig.~\ref{fig-class-deriv}) for the singular
behavior of $E(\xi,k/N,\Lambda/N)$ as the ESQPT is crossed both
``horizontally'' [$E(\xi)$] and ``vertically'' [$E(k/N)$].  Namely,
$E(\xi)$ undergoes a singularity
in which the \textit{slope} sharply
approaches zero ($\partial E/\partial\xi\<\rightarrow0$)
[Fig.~\ref{fig-class-deriv}(a)] but with a
\textit{curvature} which becomes infinite and reverses sign ($\partial^2
E/\partial\xi^2\<\rightarrow\pm\infty$), yielding a special divergent
form of inflection point, as $\xi\<\rightarrow\xicex$.  A similar
singularity is expected in $E(k/N)$ [Fig.~\ref{fig-class-deriv}(b)] at
the critical energy.

\begin{figure}
\begin{center}
\includegraphics*[width=0.8\hsize]{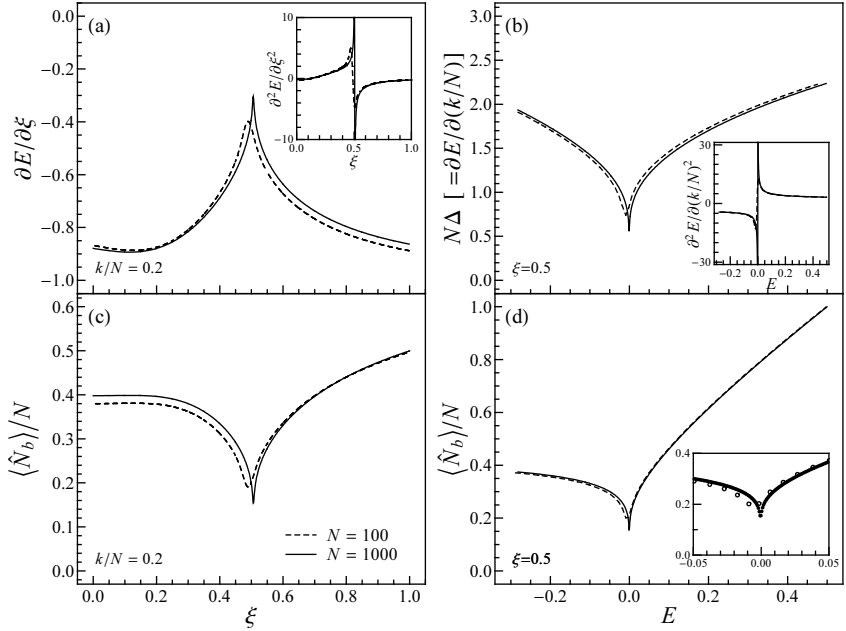}
\end{center}
\caption{Evolution of excited level energies and the order parameter
$\langle N_b \rangle$ across the ESQPT, as traversed both by varying
$\xi$~(left) and by varying $E$~(right), \textit{i.e.},
``horizontally'' and ``vertically''.  Calculations are shown for the
$\grpu{3}$ vibron model $l\<=0$ states, with $N\<=100$ (dashed curves)
and $1000$ (solid curves).  (a)~The derivatives $\partial
E/\partial\xi$ and $\partial^2 E/\partial\xi^2$ (inset), for a
specific excited level ($k/N\<=0.2$).  (b)~The derivative $\partial
E/\partial(k/N)$ or, equivalently, the scaled gap $N\Delta$, and
$\partial^2 E/\partial(k/N)^2$ (inset), for $\xi\<=0.5$. (c)~The order
parameter $\langle N_b \rangle$ (rescaled by $N$) as a function of
$\xi$ for the same level as in panel~(a).  (d)~The order parameter
$\langle N_b \rangle$ (rescaled by $N$) as a function of excitation
energy, for the same $\xi$ value as in panel~(b).  The discrete
eigenstates are resolved at the expanded scale shown in the inset.  }
\label{fig-evoln-ex-combo}
\end{figure}

The actual diagonalization results at finite $N$ show clear precursors
of this form of singularity in $E$ as $\xi$ is varied.  Even for the
small system size ($N\<=10$) considered in
Fig.~\ref{fig-evoln-models}, each eigenvalue $E(\xi)$ undergoes an
inflection [Fig.~\ref{fig-evoln-models} (solid curves)] at an energy close to the expected critical energy,
\textit{i.e.}, $\Ec=0$ for the Hamiltonians used.  The
derivative $\partial E/\partial\xi$ is shown for larger boson
number ($N\<=100$ and $1000$) in Fig.~\ref{fig-evoln-ex-combo}(a), for
the $\grpu{3}$ vibron model $l\<=0$ states.  The second derivative
$\partial^2 E/\partial\xi^2$ is also shown (inset).  The expected dip
$\partial E/\partial\xi\<\rightarrow0$ and divergent inflection
$\partial^2 E/\partial\xi^2\<\rightarrow\pm\infty$ both are present
and become gradually sharper with increasing $N$.

For nonzero $l$ in Fig.~\ref{fig-evoln-models}(b), the inflection
points in the eigenvalues as functions of $\xi$ are washed out, as
expected from the semiclassical analysis (for $l\<>0$ the centrifugal
term suppresses the probability density near $r\<=0$, mitigating the
effect of the barrier).  For $N\<=10$, the inflection points disappear
even for the very lowest nonzero $l$ values [dashed curves in
Fig.~\ref{fig-evoln-models}(b)].  Also for $N\<=10$, the inflection
points are suppressed for the negative parity ($g\<=1$) states of the
Lipkin model in Fig.~\ref{fig-evoln-models}(a).  Here a similar
mechanism applies: negative parity states posess a node at $x\<=0$,
and the effect of the parabolic barrier at $x\<=0$ is therefore again
reduced.  (To this extent, the grade in the Lipkin model is a
surrogate for the angular momentum in the higher-dimensional boson
models.  The formal relation is given in Appendix~\ref{secappboson}.)  Compare also the curves for nonzero seniorities in
Fig.~\ref{fig-evoln-models}(c,d).  While the change in behavior of the
eigenvalues between $l\<=0$ and nonzero $l$ seems to be rather abrupt
for the $N\<=10$ illustration, it must be borne in mind that the
relevant parameter for the semiclassical description was noted to be
$l/N$, which can only be varied very coarsely when $N\<=10$.  The more
gradual evolution of the ESQPT with $l/N$, as obtained for larger $N$,
is considered further below.

The properties of the spectrum as the ESQPT is traversed
``vertically'' by varying the excitation quantum number for a single
fixed Hamiltonian parameter value $\xi$ are explored in
Fig.~\ref{fig-evoln-ex-combo}(b), again for the $\grpu{3}$ vibron
model with $N\<=100$ and $1000$, now at the specific parameter value $\xi\<=0.5$.  The singularity in
$E(\xi,k/N,\Lambda/N)$ with respect to $k/N$ gives rise to the
original, defining property of the ESQPT, namely the vanishing gap or
infinite level density.  The gap is simply the change in energy for a
unit change in $k$ quantum number, so in the limit where $k/N$ is
taken as a quasi-continuous variable we have $\partial
E(\xi,k/N,\Lambda/N)/\partial (k/N) \<=N\Delta(\xi,k/N,\Lambda/N)$.
The gap is shown as a function of energy, rather than of $k$, in
Fig.~\ref{fig-evoln-ex-combo}(b), so that the energy in the spectrum
at which the precursors of the singularity occurs can be compared with
the expected critical energy $\Ec\<=0$.  The second derivative
$\partial^2 E/\partial (k/N)^2$ is also shown (inset).  The
qualitative features $\partial E/\partial (k/N)\<\rightarrow0$ and
$\partial^2 E/\partial (k/N)^2\<\rightarrow\pm\infty$ expected from
the semiclassical analysis are indeed realized, more sharply with
increasing $N$.

The inflection point of $E$ with respect to $k$ at $E\<=0$ (though not its singular nature) is also
immediately visible simply by inspection of the $l\<=0$ spectra
obtained for various $\xi$ (Fig.~\ref{fig-spectra}).  The spectra are
concave downward with respect to $k$ below $E\<=0$ and concave upward
above this energy.  At the $\grpso{3}$ limit, the entire spectrum
falls below $E\<=0$ and constant downward concavity follows from the
exact formula~\cite{iachello1996:vibron-2dim} for the eigenvalues,
quadratic in $k$.  Although here we are considering the dip in
$\partial E/\partial k$ as a property of the ESQPT in a many-body
interacting boson model, it should be noted that the dip arising for
the associated two-dimensional Schr\"odinger equation is well known as
the ``Dixon dip''~\cite{dixon1964:molecular-dip}, with applications to
molecular spectroscopy (see also Ref.~\cite{perez-bernal:u3-vibron-phase}).

For nonzero $l$, as noted above, the relevant parameter governing the
disappearance of the ESQPT is expected to be $l/N$.  The eigenvalue
spectrum for the $\grpu{3}$ vibron model with $N=100$ indeed shows
compression of the level density at the critical energy for $l\<=0$
[Fig.~\ref{fig-evoln-l}(a)] and, conversely, no apparent compression
of level density for large $l/N$ [Fig.~\ref{fig-evoln-l}(b)], where
$l\<=25$ or $l/N\<=1/4$ is shown in this example.  (See
Ref.~\cite{heinze2006:ibm-o6-u5-part1-level-dynamics} for analogous
plots for the IBM.)  However, the gradual nature of the evolution with
$l/N$ is seen by considering the dip in $\partial E/\partial k$, which
becomes continuously less deep and less sharp as $l/N$ is increased
[Fig.~\ref{fig-evoln-l}(c)].  

\subsection{Finite-size scaling}
\label{secobsscaling}

The spectroscopic hallmark of the critical point of a QPT is not a
\textit{vanishing} gap \textit{per se}, since the gap never strictly vanishes
for finite system size, but rather the \textit{nature} of its approach
to zero as $N$ increases.  It is therefore essential to characterize
the finite size scaling behavior of the gap in the vicinity of the
ESQPT.  With the Hamiltonian normalization of~(\ref{eqnHMMxi}), the
gap $\Delta$ everywhere approaches zero with increasing $N$, so we are
actually, more precisely, interested in the scaling of the gap at the
ESQPT \textit{relative} to the scaling elsewhere in the spectrum.  For
states well-separated from both the ground state QPT and ESQPT, the
scaling is as $\Delta\<\sim N^{-1}$.  For states in the vicinity of
the \textit{ground state} QPT, the gap vanishes more quickly than
$N^{-1}$, as the power law $\Delta\<\sim N^{-4/3}$.  This has been
established both numerically and analytically for the various models
under
consideration~\cite{botet1983:lipkin-scaling,rowe2004:ibm-critical-scaling,dusuel2005:lipkin-scaling,dusuel2005:bcs-scaling,dusuel2005:scaling-ibm,perez-bernal:u3-vibron-phase}.\footnote{As
noted above, different normalization conventions may be encountered for the
model Hamiltonians.  Overall multiplcation of the Hamiltonian by a
factor $N$ gives rise to a superficial difference of
unity in the finite-size scaling exponents.}  The gap at the
\textit{excited state} QPT also approaches zero more rapidly than
$\Delta\<\sim N^{-1}$.  This is apparent even from the simple plot
Fig.~\ref{fig-evoln-ex-combo}(b), where $N\Delta$ is essentially
independent of $N$ away from the critical energy (compare the curves
for $N\<=100$ and $N=1000$) but approaches zero with increasing $N$ at
the critical energy.

Let us now examine finite-size scaling more carefully, in particular, to
see the extent to which the semiclassical expression~(\ref{eqnDeltaW})
reproduces the scaling behavior.  It is not
\textit{a priori} obvious that the semiclassical
result~(\ref{eqnDeltaW}) should yield the proper scaling properties
for the eigenvalues in the vicinity of the ESQPT.  Even in the
solution of the ordinary Schr\"odinger equation, the semiclassical
analysis becomes unreliable for the first few eigenvalues in the
vicinity of the top of the barrier~\cite{ford1959:parabolic-barrier,cary1987:separatrix-eigenvalues,child1998:champagne-monodromy}.

\begin{figure}
\begin{center}
\includegraphics*[width=0.9\hsize]{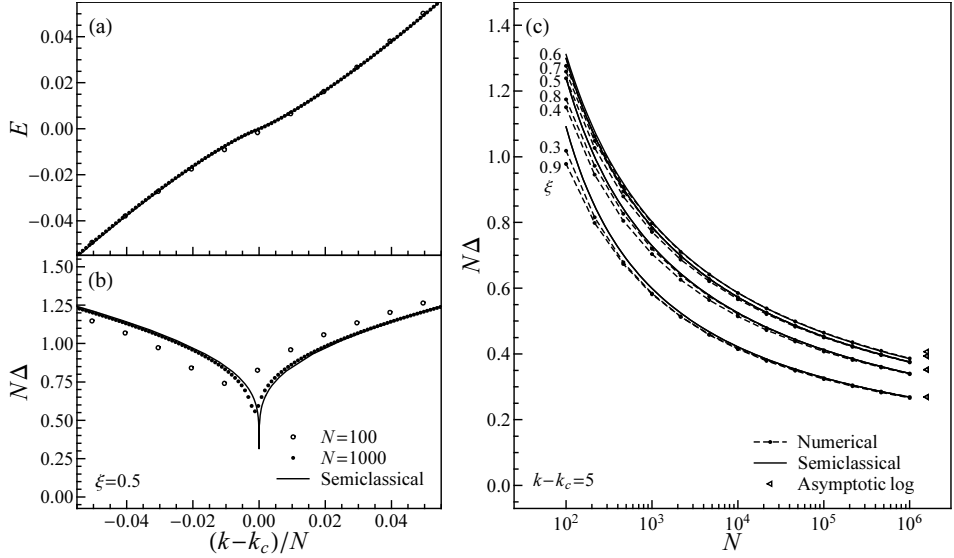}
\end{center}
\caption{Quantitative comparison of quantum and semiclassical results
for the gap, including finite-size scaling properties, in the vicinity
of the ESQPT ($E\<\approx0$).  Calculations are for the $\grpu{3}$ vibron model
$l\<=0$ states with $\xi\<=0.5$.  (a,b)~Eigenvalue spectrum and
its first difference, \textit{i.e.}, the gap, shown as functions of $(k-\kc)/N$ for
$N\<=100$ (open circles) and $1000$ (solid circles).  The
semiclassical result~(\ref{eqnEW}) or~(\ref{eqnDeltaW}) in terms of
the $W$ function (with $\alpha\<=2.49$) is shown for
comparison (solid curve). (c)~Scaling of the gap with respect to $N$,
evaluated at fixed quantum number $k-\kc\<=5$ relative to the ESQPT,
for $\xi\<=0.3$, $0.4$, $\ldots$, $0.9$.  The semiclassical results for the scaling 
(with $\alpha\<=1.24$, $1.92$, $2.24$, and $2.35$) are
shown for comparison (solid curve).  The results of the asymptotic
logarithmic expression~(\ref{eqnDeltalog}), evaluated at $N\<=10^6$, are also
indicated (open triangles).
}
\label{fig-semi-scaling}
\end{figure}

First, in Fig.~\ref{fig-semi-scaling}(a), the actual form of the
spectrum in the vicinity of the ESQPT, obtained by numerical
diagonalization, is compared with the semiclassical
estimate~(\ref{eqnEW}).  Eigenvalues are shown for $N\<=100$ and
$N\<=1000$.  Note that $\kc$ is simply determined as the value of $k$
for which the energy eigenvalues cross zero.  This must be
interpolated between discrete eigenvalues, so $\kc$ is in general
noninteger.  The singular logarithmic term in~(\ref{eqnSexpansion})
has a coefficient which is predicted unambiguously from the value of
$\hbar\omega$~(\ref{eqnhbaromega}) for the inverted oscillator, but no
attempt is made here to directly calculate the coefficient $\alpha$ of
the nonsingular linear term.  Rather, $\alpha$ is simply chosen to
numerically reproduce the linear trend in the eigenvalues in the
vicinity of the ESQPT.  The $\alpha$ value obtained from a limited
number of eigenvalues around $E\<=0$ therefore depends somewhat on
both $N$ and the number of eigenvalues considered.
The gap, that is
the first difference of the eigenvalues in
Fig.~\ref{fig-semi-scaling}(a), is plotted in
Fig.~\ref{fig-semi-scaling}(b), together with the semiclassical
estimate~(\ref{eqnDeltaW}).  The form of the singularity is well
matched by the semiclassical estimate.  (The parameter $\alpha$
essentially determines the normalization of the curve
$\Delta[(k-\kc)/N]$.)  The most significant deviation occurs for the
first few eigenvalues around $E\<=0$.

Some care must be taken in establishing exactly what gap is to be
considered in the context of finite-size scaling, since the gap is a
function of $k-\kc$, that is, how far above or below the ESQPT the gap
is measured.  The phase transition does not fall exactly ``on'' an
eigenvalue ($\kc$ is in general noninteger), the gap is varying
singularly with $k$ at $\kc$, and the quantum corrections are
fluctuating most strongly for the first few eigenvalues in the vicinity of
$\kc$~\cite{child1998:champagne-monodromy}.  Therefore, in considering
the finite-size scaling at the mean field level, it is only meaningful
to examine the gap some sufficient number of eigenvalues above or below the
ESQPT, but nonetheless close enough ($\abs{k-\kc}\<<<N$) that the scaling
appropriate to the ESQPT dominates over the usual $\Delta\<\sim
N^{-1}$ scaling.

The gap for $k-\kc\<=5$ is plotted as a function of $N$, for
$10^2\<\leq N \<\leq 10^6$, in Fig.~\ref{fig-semi-scaling}(c).  (The
quantity plotted is essentially the gap between the fifth and sixth
eigenvalues above $E\<=0$, but interpolation is necessary, since
$k-\kc$ is discrete and noninteger in the actual spectra.)  Note
foremost that the gaps for $\xi\<=0.3$ and $\xi\<=0.9$, or for
$\xi\<=0.4$ and $\xi\<=0.8$, or for $\xi\<=0.5$ and $\xi\<=0.7$,
converge towards each other for large $N$.  Since
$\Xi(\xi)$ is symmetric about $\xi\<=0.6$ [see~(\ref{eqnXi})], this
demonstrates that the asymptotic behavior depends on $\xi$ through
$\Xi(\xi)$, as expected if the properties of the ESQPT are dominated
by the $\hbar\omega$ value~(\ref{eqnhbaromega}) of the parabolic top
of the barrier.  The semiclassical estimate~(\ref{eqnDeltaW}) is shown
for comparison (using only one fixed $\alpha$ value for each symmetric
pair of $\xi$ values, for simplicity) and appears to reasonably
reproduce the finite-size scaling.  The results of the simple
logarithmic approximation $N\Delta\<\approx 2\pi \Xi(\xi)^{1/2}/N$
from~(\ref{eqnDeltalog}), evaluated at $N\<=10^6$, are also shown for
reference.

\subsection{Order parameters}
\label{secobsorder}

Let us now consider the singularity in the order parameter $\langle
\Nhat_b \rangle$ (or $\langle \Nhat_2 \rangle$), which plays a defining role
for the ground state QPT.  The evolution of the order parameter
$\langle \Nhat_b \rangle_k$ is shown as a function of $\xi$ in
Fig.~\ref{fig-evoln-ex-combo}(c), again for the $\grpu{3}$ vibron
model, for the same level ($k/N\<=0.2$) considered in
Fig.~\ref{fig-evoln-ex-combo}(a).  This quantity is closely related to
the energy plotted in Fig.~\ref{fig-evoln-ex-combo}(a), since
\begin{equation}
\frac{dE_k(\xi)}{d\xi}
=\frac{1}{\xi}\Bigl[E_k(\xi)-\frac{\langle \Nhat_b\rangle_k}{N}\Bigr]
\end{equation}
by the Feynman-Hellmann theorem.

It is seen that $\langle \Nhat_b \rangle_k$ undergoes a dip towards zero
at $\xi\<=\xicex$, which becomes sharper and deeper with increasing
$N$.  At the semiclassical level, one of the essential characteristics
of the ESQPT was localization of the wave function at $x\<=0$,
together with vanishing classical velocity (hence, $p^2\<=0$).  In
coordinate form, $\Nhat_b\<\propto p^2/N^2 + x^2$ [with the coordinate
definitions used in~(\ref{eqnHclassx})], so the natural extension to
the fully quantum description is localization of probability with
respect to occupation number at $N_b\approx0$.  The order parameter is
shown as a function of energy in Fig.~\ref{fig-evoln-ex-combo}(d), for
the same fixed $\xi$ value ($\xi\<=0.5$) considered in
Fig.~\ref{fig-evoln-ex-combo}(b).  The ``evolution'' of properties
with respect to excitation energy is of necessity discrete, since for
finite $N$ the eigenvalue spectrum is itself discrete
[Fig.~\ref{fig-evoln-ex-combo}(d) inset].
It is apparent from Fig.~\ref{fig-evoln-ex-combo}(c,d) that, while
$\langle \Nhat_b \rangle_k$ drops \textit{towards} zero at the ESQPT, and
the dip becomes sharper and deeper with increasing $N$, $\langle \Nhat_b \rangle_k$ is far from
actually reaching zero at the finite $N$ being considered.

\section{Quantum phases}
\label{secphase}

So far we have considered the excited state quantum phase transition
as a \textit{singularity} in the \textit{evolution} of the excited
state properties rather than as a \textit{boundary} between
\textit{phases}.  A central question which arises in connection with the ESQPT
phenomenology concerns the meaning of ``phases'' for excited states,
namely, whether or not the excited states on each side of the phase
transition can meaningfully be considered to belong to qualitatively
distinct phases.  Of course, in thermodynamics, it is well known that
\textit{phase transitions}, in the sense of singularities, 
do not necessarily imply the existence of distinguishable
\textit{phases}, the liquid-vapor transition in the vicinity of a
critical point being a classic counterexample.  Here we approach
identification of phases both through indirect measures of the
structural properties of the states on either side of the ESQPT
(\textit{e.g.}, order parameters and spectroscopic signatures) and
directly through inspection of the wave functions.

For the ground state, the ``phase'' is simply indicated by the value
of the order parameter $\langle \Nhat_b\rangle_0$ (or $\langle
\Nhat_2\rangle_0$).  In the large $N$ limit,
the value of $\langle \Nhat_b\rangle_0$ is qualitatively different on
either side of $\xic$, namely, vanishing for $\xi\<<\xic$ and nonzero
(growing towards $N/2$) for $\xi\<>\xic$.  In contrast, for the
excited states, $\langle \Nhat_b \rangle_k$ does not show such a
qualitative difference between the two sides of the ESQPT.  Rather,
$\langle \Nhat_b\rangle_k\<\rightarrow0$ as the level $k$ crosses the
ESQPT but is nonzero on \textit{either} side (Sec.~\ref{secobsorder}).
Therefore, the expectation value $\langle \Nhat_b\rangle_k$ by itself
does not distinguish two ``phases'' for the excited states.

The reason is fundamentally related to the classical limit of the
problem (Sec.~\ref{secclass}).  Recall that $\Nhat_b\<\propto p^2/N^2
+ x^2$.  For the classical ground state, the kinetic energy vanishes,
and the static equilibrium value for $x$ is simply determined by the
location of the minimum in the potential~(\ref{eqnVclass}).  For
excited states, such a \textit{static} quantity no longer provides a suitable
measure of the phase at the classical level, since excited states (with
nonzero kinetic energy) are not described by a single equilibrium
position.  Instead, one must consider a \textit{dynamical} definition
of phase, taking into account the topology of the classical orbits in
the phase
space~\cite{macek2006:ibm-o6-u5-part2-classical-trajectories,cejnar2006:excited-qpt}.
The classical analogue of the ``expectation value'' of an observable
is its time average over the classical motion, $\langle f(x,p)
\rangle \<\equiv \tau^{-1} \oint f(x,p) v(E,x)^{-1}\,dx$.
This is also the semiclassical average with respect to the first-order
WKB probability density $P(x)\<\propto v(E,x)^{-1}$, so the time
average carries over naturally to the quantum expectation value.  At
the quantum level, the consequence of the breakdown of the static
definition is that the expectation value $\langle \Nhat_b\rangle_k$ does
not provide an unambiguous measure of the phase of an excited state.
\begin{figure}
\begin{center}
\includegraphics*[width=0.7\hsize]{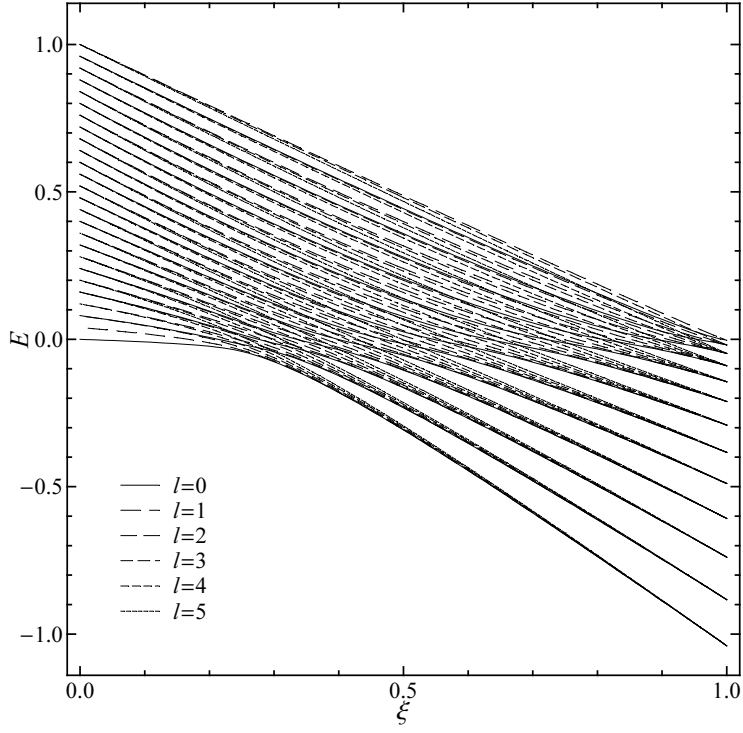}
\end{center}
\caption{
Correlation diagram for the $\grpu{3}$ vibron model ($N\<=25$), with
$0\<\leq l \<\leq 5$, showing the change in angular momentum
degeneracies across the ESQPT.
}
\label{fig-correlation}
\end{figure}
\begin{figure}
\begin{center}
\includegraphics*[width=0.9\hsize]{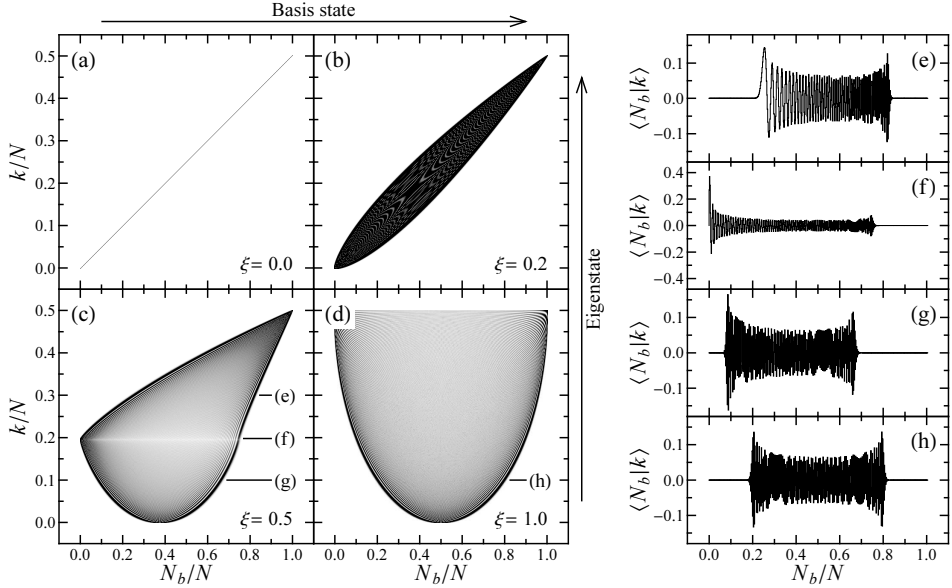}
\end{center}
\caption{
Probability distributions for the entire spectrum of eigenstates,
decomposed with respect to the $N_b$ quantum number, \textit{i.e.}, in
the $\grpu{2}$ basis, for the $\grpu{3}$ vibron model with $l\<=0$ and
$N\<=1000$.  The probability distributions are shown for
(a)~$\xi\<=0$, (b)~$\xi\<=0.2$, (c)~$\xi\<=0.5$, and (d)~$\xi\<=1$.
Also shown are the wave functions for individual representative
states: (e)~above the ESQPT ($\xi\<=0.5$, $k/N\<=0.3$), (f)~at the
ESQPT ($\xi\<=0.5$, $k/N\<=0.2$), (g)~below the ESQPT ($\xi\<=0.5$,
$k/N\<=0.1$), and (h)~for the $\grpso{3}$ dynamical symmetry
($\xi\<=1$, $k/N\<=0.1$).  }
\label{fig-eigenprob}
\end{figure}

Nonetheless, there are qualitative changes in the spectrum across the
ESQPT.  In particular, the degeneracy patterns with respect to the
angular momentum (or seniority) quantum number change from those
characteristic of the $\grpu{n_1}\otimes\grpu{n_2}$ dynamical symmetry
above the critical energy to those characteristic of the
$\grpso{n_1+n_2}$ dynamical symmetry below the critical energy.  At
the critical energy, a rapid rearrangement of degeneracies occurs.
This is clearly visible for all the models in
Fig.~\ref{fig-evoln-models}.  The evolution of the eigenvalues for the
$\grpu{3}$ model is shown in detail in Fig.~\ref{fig-correlation}, for
more angular momentum values ($l\<\leq5$) and for a larger particle
number ($N\<=25$) than in Fig.~\ref{fig-evoln-models}(b).  Note that
the like-parity states ($l$ odd or $l$ even) form approximate
degenerate multiplets [$\grpu{2}$-like] above $E\<=0$ for all $\xi$,
while multiplets are composed of all $l$ values [$\grpso{3}$-like]
below this energy.  For the Lipkin model, the transition between
degeneracy patterns is understood from the geometric Hamiltonian, as
noted in Ref.~\cite{heiss2002:qpt-instability}, in terms of degenerate
parity doublets below the barrier and lifting of this degeneracy above
the barrier.  For the higher-dimensional models, the change in degeneracies
at the ESQPT is indicative of the breakdown of the adiabatic seperation of
rotational and radial vibration degrees of freedom at the critical energy.

For the ground state QPT, the persistence of the degeneracies
associated with the symmetry limits as the QPT is approached, in spite
of strong symmetry-breaking interactions, has been explained in terms
of quasidynamical symmetry~\cite{carvalho1986:sp-shell-collective}.
The qualitative distinction between the states on either side of the
QPT lies in their forming approximate embedded representations of
either the $\grpu{n_1}\otimes\grpu{n_2}$ or $\grpso{n_1+n_2}$
algebras.  (In particular, the phases obtained on either side of the
ground state QPT have been characterized for the IBM in
Ref.~\cite{rowe2004:quasidynamical-1-ibm-u5-so6}.)  We therefore note
that it is of considerable interest to determine whether or not there
is a similar sharp distinction between the states, as forming
approximate embedded representations of one or the other of these
algebras, across the ESQPT.

To consider the question of phases further, let us inspect the
structure of the wave functions for the $\grpu{3}$ vibron model
excited states, as decomposed in the $\grpu{2}$ (good $N_b$) dynamical
symmetry basis.  Each density plots in Fig.~\ref{fig-eigenprob}(a--d)
concisely sumarizes the decomposition the entire spectrum of $l\<=0$
eigenstates, for a given value of $\xi$.  [A horizontal slice
across the plot gives the ``wave function'' of one excited state or,
more precisely, the squared amplitudes in its decomposition with
respect to the $\grpu{2}$ basis.  The ground state is represented by
the bottommost slice.]

To provide context, first consider the structure of the states when no
ESQPT is present.  For $\xi\<=0$, the Hamiltonian is diagonal in the
$\grpu{2}$ basis [Fig.~\ref{fig-eigenprob}(a)].  As $\xi$ increases
towards $0.2$, the ground state critical value, there is a spreading
of the probability distribution over many neighboring basis states,
essentially confined to a teardrop shaped region of the $N_b$-$k$ plot
[Fig.~\ref{fig-eigenprob}(b)].  At the other limit, $\xi\<=1$, where
the $\grpso{3}$ dynamical symmetry occurs, the probability
decomposition also follows a regular pattern
[Fig.~\ref{fig-eigenprob}(d)].  Here the wave functions for the
eigenstates with respect to the $\grpu{2}$ basis are known
analytically~\cite{santopinto1996:so-brackets,perez-bernal:u3-vibron-phase}.
The probability distribution for each state is reflection symmetric
about $N_b/N\<=0.5$, peaked at two symmetric extreme values.

The relevant plot for consideration of the ESQPT is now
Fig.~\ref{fig-eigenprob}(c), where the probability decompositions of
the eigenstates are shown for $\xi\<=0.5$.  Individual wave functions
are shown in Fig.~\ref{fig-eigenprob}(e--g), with an $\grpso{3}$ wave function
[Fig.~\ref{fig-eigenprob}(h)] given for comparison.  Below the critical
energy ($k/N\<\approx0.2$), the probability decompositions, taken in
aggregate, bear a marked resemblance to those obtained for the
$\grpso{3}$ dynamical symmetry in Fig.~\ref{fig-eigenprob}(d).
However, they are scaled towards smaller $N_b$ and, in particular, are
approximately reflection symmetric about a reduced value of $N_b$
($N_b/N\<\approx0.4$).  In the immediate vicinity of the critical
energy, the probability is moderately localized at low $N_b$.  Above the critical
energy, the probability distribution for each eigenstate is again
strongly peaked at two extreme values of $N_b$, but these values are
not symmetric about a fixed $N_b$ as they are below the critical
energy.  Rather, their midpoint increases approximately linearly with
excitation quantum number, as reflected in the linear behavior of
$\langle \Nhat_b \rangle$ above $E\<=0$ in
Fig.~\ref{fig-evoln-ex-combo}(b).

The qualitative distinction between the wave functions below and above
the critical energy is therefore clearly apparent when the states are
viewed in \textit{aggregate}, as in Fig.~\ref{fig-eigenprob}(c).  For
any given interaction parameter value $\xi$, the qualitative
distinction also apparently involves reflection symmetry (or lack
thereof) about some fixed $N_b/N\<<0.5$.  However, the appropriate
means of constructing a simple measure which allows the immediate
characterization of the ``phase'' of a state taken in
\textit{isolation} is not obvious and requires further
consideration.

%%%%%%%%%%%%%%%%%%%%%%%%%%%%%%%%%%%%%%%%%%%%%%%%%%%%%%%%%%%%%%%%
%%%%%%%%%%%%%%%%%%%%%%%%%%%%%%%%%%%%%%%%%%%%%%%%%%%%%%%%%%%%%%%%

\section{Conclusions and outlook}
\label{secconcl}

In the present work, we have seen that the ESQPT phenomena are
universal to a broad family of two-level models with pairing
interactions, including not only the $s$-$b$ models (\textit{e.g.},
Lipkin model, vibron model, and IBM) but also the generic two-level
bosonic and fermionic pairing models.  The properties of the
eigenvalue spectra (including the quantum gap or level density) and
order parameters in the vicinity of the ESQPT have been investigated,
both at the semiclassical level and numerically for the full quantum
problem.  The finite-size scaling properties of the gap have been
considered at the mean-field level.  Qualitative differences between
distinct ``phases'' on either side of the ESQPT have also been
noted.  In the process, both parallels with and differences from
the conventional ground state QPT have been identified.

An important aspect is the experimental evidence for ESQPTs.  This
requires the identification of physical systems described by algebraic
Hamiltonians with $\xi\<>\xic$ for which states with quantum numbers
$k\<>>1$ can be observed.  The most promising situations thus far are in
molecules described by $s$-$b$ boson
models~\cite{perez-bernal:u3-vibron-phase}.  Further examples are
needed to fully understand the experimental implications of the ESQPT.

At the theoretical level, several interesting questions remain even
for the basic two-level models considered so far. Here the main intent
was to note the aspects of the ESQPT common to the full family of
two-level models.  A detailed investigation of the specific properties
of the bosonic and fermionic two-level models with nontrivial
degeneracies for both levels is in order~\cite{caprio:two-level-alg}.
In particular, an investigation of the
$\grpu{n_1+n_2}/[\grpu{n_1}\otimes\grpu{n_2}]$ geometry of the models
is necessary.  While the coherent state analysis has been pursued
extensively for the $s$-$b$
models~\cite{dieperink1980:ibm-phase,ginocchio1980:ibm-coherent-bohr,feng1981:ibm-phase,hatch1982:ibm-classical,leviatan1987:ibm-intrinsic},
only preliminary use of coherent states has so far been made for the generic
pairing models~\cite{chen1995:pair-coupling}.

Also, the role of integrability~\cite{tabor1989:chaos-integrability}
in the ESQPT must be explored.  The $\grpso{n}$-invariant
Hamiltonian~(\ref{eqnHMMxi}), or more generally the
$\grpso{n_1}\otimes\grpso{n_2}$-invariant or
$\grpsp{n_1}\otimes\grpsp{n_2}$-invariant
Hamiltonian~(\ref{eqnHPPxi}), is integrable and, moreover, leads to a
separable and effectively one-dimensional problem~(\ref{eqnHclassr}) in the
classical limit.  Application of the quantization
condition~(\ref{eqnWKBquant}), which played a crucial role in the
semiclassical analysis, is limited to integrable (or approximately
integrable) systems.  However, interactions beyond the pure pairing
interaction are necessary for realistic applications.  These
interactions destroy integrability.  They also in general give rise to
\textit{first-order} ground state QPTs, for which even the ground
state scaling properties have only been partially
characterized~\cite{iachello2004:qpt-mesoscopic,rowe2004:ibm-critical-scaling}.
Possible manifestation of an ESQPT in the spectrum of a nonintegrable
Hamiltonian is discussed in
Ref.~\cite{macek2007:ibm-chaos-regularity}, but a general theoretical
foundation for ESQPTs in nonintegrable systems has yet to be
developed.

It would be valuable to bridge the gap between the ground state QPT,
where the quantum properties scale according to power laws, and the
ESQPT, where the singularity is logarithmic in nature.
Semiclassically, the connection between the two is nontrivial, since
the ground state QPT involves a pure quartic potential (\textit{i.e.},
no barrier) [Fig.~\ref{fig-class-action}(b)], while the ESQPT analysis
[Fig.~\ref{fig-class-action}(c)] requires the classical turning point
at the barrier to be well-separated from the classical turning point
at the outer wall of the well [\textit{i.e.}, a high barrier or,
conversely, small $\hbarfracinline\sim 1/N$], so that many states
lie below the ESQPT.  In actual spectroscopic applications,
such as to nuclei, often only relatively few low-lying states can be
observed experimentally.  Therefore, the finite-$N$ precursors of the
ESQPT in the intermediate regime, where the ESQPT is low-lying in the
spectrum, are of special interest.

The main physical interest, however, lies in possible broad relevance
of the ESQPT phenomena to various mesoscopic systems, at least those
dominated by pairing interactions.  In this regard, the analysis must
be extended to more realistic multi-level pairing models.  For
instance, the multi-level pairing model with equally-spaced levels is
of special interest for application to the spectra of superconducting
grains~\cite{vondelft2001:supercond-grain-spectroscopy}.  Multi-level
pairing models can also provide a foundation for realistic
calculations with the nuclear shell
model~\cite{volya2001:pairing-quasispin}.  The models considered in
the present work may be constructed as the ``infinitely-coordinated''
limit~\cite{botet1983:lipkin-scaling} of the Ising-type spin-lattice
models, \textit{i.e.}, the limit in which all sites interact equally
with all others by a long-range interaction.  It would thus be of
interest to examine under what conditions an ESQPT may occur in such
models for finite-range interactions.  (It has recently been
shown~\cite{pan-PREPRINT:pairing-boson-mapping-hubbard} that the
fermionic two-level pairing Hamiltonian is related to a Bose-Hubbard
Hamiltonian~\cite{hubbard1963:electron-correlations} by an exact boson
mapping, suggesting possible application of the ESQPT concept to
correlated electron systems or to ultracold atoms trapped in optical
lattices.)  An alternate avenue for extension to realistic systems is
through coupling of multiple two-level systems as subsystems,
\textit{e.g.}, the Dicke model~\cite{dicke1953:model-superradiance}
for quantum optical systems may be obtained as two coupled Lipkin
models.

\begin{ack}
Discussions with F.~P\'erez-Bernal and S.~Frauendorf are gratefully
acknowledged.  This work was supported by the US Department of Energy
(grant DE-FG02-91ER-40608), the Czech Science Foundation
(project 202/06/0363), and the Czech Ministry of Education, Youth, and
Sports (project MSM 0021620859) and was carried out in part at the
European Centre for Theoretical Studies in Nuclear Physics and Related
Areas (ECT*).
\end{ack}

%%%%%%%%%%%%%%%%%%%%%%%%%%%%%%%%%%%%%%%%%%%%%%%%%%%%%%%%%%%%%%%%
%%%%%%%%%%%%%%%%%%%%%%%%%%%%%%%%%%%%%%%%%%%%%%%%%%%%%%%%%%%%%%%%

\appendix

\section{Quasispin and multipole Hamiltonians for the 
two-level boson model}
\label{secappboson}

The two-level boson model [Fig.~\ref{fig-levels}(a)] is characterized
by two overlaid algebraic structures: a $\grpu{n+1}$ structure ($n\<=2L+1$)
arising from the bilinears in the creation and annihilation operators
and an $\grpsu{1,1}$ structure involving the pairing quasispin
operators.  The twin algebraic structures provide a simple
relationship between the multipole Hamiltonian~(\ref{eqnHMMxi}) and
pairing Hamiltonian~(\ref{eqnHquasi}) for the system.  In this
appendix, we summarize the relevant algebraic properties and deduce
the explicit relationship between the pairing and multipole
Hamiltonians, for arbitrary $L$ and for the both possible phase
choices.  The relationship noted for the IBM in
Ref.~\cite{arima1979:ibm-o6} is recovered as a special case.

The well-known bosonic $\grpsu{1,1}$ quasispin
algebra~\cite{ui1968:su11-quasispin-shell} or fermionic $\grpsu{2}$
quasispin algebra~\cite{kerman1961:pairing-collective} generators are
given, in the convention we adopt here, by
\begin{equation}
\label{eqnSdefn}
\Shat_{j+}\equiv \tfrac12\sum_m c_{jm}^\dagger \tilde{c}_{jm}^\dagger
\quad
\Shat_{j-}\equiv \tfrac12 \sum_m \tilde{c}_{jm} c_{jm}
\quad
\Shat_{jz}\equiv \tfrac14 \sum_m
(c_{jm}^\dagger \tilde{c}_{jm} +
\tilde{c}_{jm} c_{jm}^\dagger)
\end{equation}
and obey commutation relations
\begin{equation}
\label{eqnScomm}
[\Shat_{j+},\Shat_{j-}]=\mp2\Shat_{jz} \quad [\Shat_{jz},\Shat_{j+}]=+\Shat_{j+} \quad [\Shat_{jz},\Shat_{j-}]=-\Shat_{j-},
\end{equation}
where the upper and lower signs apply to the bosonic and
fermionic cases, respectively.
The quasispin $z$ projection $\Shat_{jz}$ is
simply related to the occupancy $\Nhat_j\<\equiv\sum_m c_{jm}^\dagger
c_{jm}$, by
$\Shat_{jz}=\tfrac12(\Nhat_j\pm\Omega_j)$, where $\Omega_j$ is the
half-degeneracy of level $j$.
The squared quasispin 
\begin{equation}
\label{eqnSsqr}
\Svechat_j^2
\<\equiv \Shat_{jz}^2\mp\tfrac12(\Shat_{j+}\Shat_{j-}+\Shat_{j-}\Shat_{j+})
\<= \Shat_{jz}(\Shat_{jz}-1)\mp\Shat_{j+}\Shat_{j-},
\end{equation}
with eigenvalues $\langle\Svechat_j^2\rangle\<=S_j(S_j\mp1)$, is
conserved separately for each level by the Hamiltonian~(\ref{eqnHquasi}).
Eigenstates are therefore characterized by seniority quantum numbers
$v_j\<=0,1,\ldots$, defined by $S_j\<=\tfrac12(\Omega_j\pm v_j)$.

For the two-level bosonic system defined in terms of a singlet level
$s^{(0)}$ and an $n$-fold degenerate level $b^{(L)}$ ($n\<=2L+1$), the
quasispin generators~(\ref{eqnSdefn}) are given
explicitly, in tensor notation, by
\begin{equation}
\label{eqnalgsb}
\begin{aligned}
\Shat_{s+}&=\tfrac12 (s^\dagger \cdot s^\dagger) & \Shat_{b+}&=\tfrac12 (-)^L (b^\dagger \cdot b^\dagger) \\
\Shat_{s-}&=\tfrac12 (\tilde{s} \cdot \tilde{s}) & \Shat_{b-}&=\tfrac12 (-)^L (\tilde{b} \cdot \tilde{b}) \\
\Shat_{sz}&=\tfrac14 (s^\dagger \cdot \tilde{s} + \tilde{s} \cdot
s^\dagger) & \Shat_{bz}&=\tfrac14 (-)^L (b^\dagger \cdot \tilde{b} +
\tilde{b} \cdot b^\dagger)\\
&=\tfrac12 (\Nhat_s+\tfrac12) &&=\tfrac12 (\Nhat_b+L+\tfrac12) ,
\end{aligned}
\end{equation}
where $\tilde{T}^{(\lambda)}_\mu\<\equiv(-)^{\lambda-\mu}T^{(\lambda)}_{-\mu}$
and $U^{(\lambda)}\cdot V^{(\lambda)}\<\equiv(-)^L
(2L+1)^{1/2}(A\times B)^{(0)}\<=\sum_\mu(-)^{\mu}U^{(\lambda)}_{\mu}V^{(\lambda)}_{-\mu}$.
The generators of the $\grpsu[s]{1,1}$ and $\grpsu[b]{1,1}$ algebras
can be combined to form a sum-quasispin algebra with two possible
relative phases, yielding $\grpsu[\pm]{1,1}$ algebras with generators
\begin{equation}
\label{eqnsu11pm}
\grpsu[\pm]{1,1}: \quad \Shat_+=\Shat_{s+}\pm \Shat_{b+} \quad \Shat_-=\Shat_{s-}\pm
\Shat_{b-} \quad \Shat_z=\Shat_{sz}+ \Shat_{bz} .
\end{equation}
The subalgebras of $\grpsu[s]{1,1}\otimes\grpsu[b]{1,1}$, and their
associated quantum numbers, are
\begin{equation}
\label{eqnchainsu11}
\renewcommand{\arraystretch}{1.3}
\gqn{\grpsu[s]{1,1}}{v_s=0,~1}\otimes\gqn{\grpsu[b]{1,1}}{v_b}
\supset
\left\lbrace
\begin{array}{c}
\gqn{\grpsu[+]{1,1}}{v_+}\\
\gqn{\grpu[s]{1}}{N_s}\otimes\gqn{\grpu[b]{1}}{N_b}\\
\gqn{\grpsu[-]{1,1}}{v_-}.
\end{array}
\right.
\end{equation}
The $\grpsu[s]{1,1}$ algebra for the singlet $s$-boson level is
trivial, in that $\Svechat^2\<=-3/4$ identically by application of the
canonical commutation relations for $s^\dagger_0$ and $s_0$.  This
constrains $v_s$ to the values $0$ or $1$.  Since $S_z-S$ is integral
for $\grpsu{1,1}$ representations, it follows that $v_s\<=0$
for $N_s$ even and $v_s\<=1$ for $N_s$ odd (\textit{i.e.}, $v_s\<\cong
N_s\mod{2}$).

At fixed total particle number $N$ (and therefore fixed
$S_z$), the operator $\Shat_+ \Shat_-$ is trivially related to the
Casimir invariant $\Svechat^2$ of $\grpsu[\pm]{1,1}$
by~(\ref{eqnSsqr}), as $\Shat_+\Shat_-\<=S_z(S_z-1)-\Svechat^2$, with
eigenvalues
\begin{math}
\tfrac14[N(N+2L)-v_\pm(v_\pm+2L)]
\end{math}.  
A pairing Hamiltonian~(\ref{eqnHquasi}) with pairing interaction
chosen proportional to the $\grpsu[\pm]{1,1}$ Casimir operator is thus
given by
\begin{equation}
\label{eqnHPP}
(\Hhat_{PP})_\pm=\varepsilon \Nhat_b + 4\kappa (-)^{L+1}(\Shat_{s+}\pm
\Shat_{b+})(\Shat_{s-}\pm \Shat_{b-}),
\end{equation}
where the coefficient on the last term is chosen for convenience
below.

The two-level boson system is alternatively characterized by the Lie
algebra $\grpu{n+1}$, with tensor-coupled generators
\begin{equation}
\label{eqnsun1}
\grpu{n+1}: \quad (s^\dagger\times\tilde{s})^{(0)} 
\quad (s^\dagger\times\tilde{b})^{(L)}
\quad (b^\dagger\times\tilde{s})^{(L)}
\quad (b^\dagger\times\tilde{b})^{(\lambda)}
,
\end{equation}
for $\lambda\<=0,1,\ldots,2L$.  Two distinct $\grpso[\pm]{n+1}$
subalgebras are obtained, with generators
\begin{equation}
\label{eqnson1}
\grpso[\pm]{n+1}: \quad 
\left\lbrace \begin{array}{l}
(s^\dagger\times\tilde{b})^{(L)} + (b^\dagger\times\tilde{s})^{(L)}
\\
i[ (s^\dagger\times\tilde{b})^{(L)} - (b^\dagger\times\tilde{s})^{(L)}
]
\end{array} \right\rbrace
\quad (b^\dagger\times\tilde{b})^{(\lambda)}
,
\end{equation}
for $\lambda$ restricted to odd values.  In the
case of the IBM, it is the $\grpso[+]{6}$ algebra which contains the
physical quadrupole operator~\cite{vanisacker1985:ibm-so6}.
[Therefore, conventionally, the $\grpso[+]{6}$ algebra is simply
denoted by $\grpso{6}$, while the alternate $\grpso[-]{6}$ algebra is
denoted by $\overline{\grpso{6}}$.]  An $n$-dimensional rotation
algebra $\grpso{n}$ is obtained by retaining only the generators
$(b^\dagger\times\tilde{b})^{(\lambda)}$ ($\lambda$ odd), and an
$\grpso{3}$ algebra by retaining only
$(b^\dagger\times\tilde{b})^{(1)}$~\cite{iachello2006:liealg}.

The subalgebras of $\grpu{n+1}$, and their associated quantum numbers,
are thus
\begin{equation}
\label{eqnchainun}
\renewcommand{\arraystretch}{1.3}
\gqn{\grpu{n+1}}{N}
\supset
\left\lbrace
\begin{array}{c}
\gqn{\grpso[+]{n+1}}{\sigma_+}\\
\gqn{\grpu[s]{1}}{N_s}\otimes
\gqn{\grpu{n}}{N_b}\\
\gqn{\grpso[-]{n+1}}{\sigma_-}
\end{array}
\right\rbrace
\supset
\gqn{\grpso{n}}{v}
\supset
\gqn{\grpso{3}}{J} .
\end{equation}
The Casimir operators of the subalgebras are, explicitly,
\begin{equation}
\label{eqncasimir}
% symmetric negative hspace is to tease LaTeX into allowing equation
% number alongside equation in whitespace underneath long top line,
% without affecting the actual centering of the equation body
\hspace{-1in}
\begin{aligned}
C_2[\grpso[\pm]{n+1}]&=(\pm)2(s^\dagger\times\tilde{b}\pm
b^\dagger\times\tilde{s})^{(L)}
\cdot
(s^\dagger\times\tilde{b}\pm
b^\dagger\times\tilde{s})^{(L)}+C_2[\grpso{n}]
\\
C_2[\grpso{n}]&=4\sum_{\lambda~\text{odd}}
(b^\dagger\times\tilde{b})^{(\lambda)}\cdot(b^\dagger\times\tilde{b})^{(\lambda)}
\\
C_1[\grpu{n}]&=(-)^{L}(b^\dagger\cdot\tilde{b})=\Nhat_b
\\
C_2[\grpu{n}]&=\sum_\lambda
(b^\dagger\times\tilde{b})^{(\lambda)}\cdot(b^\dagger\times\tilde{b})^{(\lambda)},
\end{aligned}
\hspace{-1in}
\end{equation}
with eigenvalues $2\sigma_\pm(\sigma_\pm+2L)$, $2v(v+2L-1)$, $N_b$ and
$N_b(N_b+2L)$, respectively.  The mathematically natural Casimir form
of the Hamiltonian is 
\begin{equation}
(\Hhat_{C})_\pm=\varepsilon N_b + \frac{\kappa}{2}
C_2[\grpso[\pm]{n+1}].
\end{equation}
The two phase choices for $\grpso[\pm]{n+1}$
yield identical eigenvalue spectra but different
eigenstates~\cite{vanisacker1985:ibm-so6}.  For physical reasons, the
$\grpso{n}$-invariant ``multipole-multipole'' Hamiltonian
\begin{equation}
\label{eqnHMM}
(\Hhat_{MM})_\pm=\varepsilon \Nhat_b + \kappa
(\pm)[(s^\dagger\times\tilde{b}\pm b^\dagger\times\tilde{s})^{(L)}
\cdot
(s^\dagger\times\tilde{b}\pm b^\dagger\times\tilde{s})^{(L)} ]
\end{equation}
is commonly used.  At fixed $v$, $\Hhat_C$ and $\Hhat_{MM}$ differ only by a
constant offset, with 
\begin{math}
(\Hhat_{MM})_\pm \<= (\Hhat_{C})_\pm - (\kappa/2) C_2[\grpso{n}]
\end{math},
by~(\ref{eqncasimir}). 

To relate the $\grpu{n+1}$ and $\grpsu{1,1}$ descriptions, let us
first observe that the Casimir operators of $\grpso{n}$ and
$\grpsu[b]{1,1}$ are related, and that the $\grpso{n}$ angular
momentum quantum number $v$ and the $\grpsu[b]{1,1}$ seniority $v_b$
are actually identical.  This is an example of a general correspondence
between the algebras $\grpso{n}$ and
$\grpsu{1,1}$~\cite{pan1988:son-isf-brackets,rowe2005:radial-me-su11}.
According to the basic quasispin relations above,
\begin{equation}
\label{eqnssqrspsm}
4\Svechat^2_b = (\Nhat_b+L+\tfrac12)(\Nhat_b+L-\tfrac32)-4\Shat_{b+}\Shat_{b-},
\end{equation}
and $4\Shat_{b+}\Shat_{b-}\<=(b^\dagger\cdot
b^\dagger)(\tilde{b}\cdot\tilde{b})$ can be recast in terms of coupled
bilinears in the creation and annihilation operators (\textit{e.g.},
by the tensor contraction methods of
Ref.~\cite{chen1993:wick-coupled}) as
\begin{equation}
\label{eqnspsmmulti}
4\Shat_{b+}\Shat_{b-} =(-)^{L+1} (b^\dagger\cdot\tilde{b}) +
\sum_\lambda (-)^\lambda
(b^\dagger\times\tilde{b})^{(\lambda)}\cdot(b^\dagger\times\tilde{b})^{(\lambda)}
.
\end{equation}
The right hand side we reexpress in terms of the Casimir
operators~(\ref{eqncasimir}) as
\begin{equation}
\label{eqnspsmcasimir}
4\Shat_{b+}\Shat_{b-} = -\Nhat_b+C_2[\grpu{n}]-\tfrac12C_2[\grpso{n}]
.
\end{equation}
On the other hand, 
\(
\langle4\Svechat^2\rangle
\<=
(v_b+2L+\tfrac12)(v_b+2L-\tfrac32)
\)
from the definition of seniority.  Comparison with the result for
$\langle4\Svechat^2\rangle$ obtained from~(\ref{eqnssqrspsm})
and~(\ref{eqnspsmcasimir}) establishes the identity of $v$ and $v_b$.

Comparison of the full Hamiltonians $\Hhat_{PP}$~(\ref{eqnHPP}) and
$\Hhat_{MM}$~(\ref{eqnHMM}) is then straightforward, with the aid
of~(\ref{eqnspsmcasimir}), yielding
\begin{equation}
\label{eqnHdiff}
(\Hhat_{MM})_\pm = (\Hhat_{PP})_{[\mp (-)^L]} + \kappa (-)^L [N(N+2L) -
v(v+2L-1)] .
\end{equation}
Thus, for $L$ even, the Hamiltonians $(\Hhat_{MM})_\pm$ and
$(\Hhat_{PP})_{\mp}$ differ only by a constant (\textit{i.e.}, a function
of conserved quantum numbers), while, for $L$ odd, it is
$(\Hhat_{MM})_\pm$ and $(\Hhat_{PP})_{\pm}$ which differ only by a constant.
Observe, therefore, that for the $\grpu{2}$, $\grpu{6}$,
\textit{etc.}, models ($L$ even), the conventional $\grpsu[+]{1,1}$
phase choice for the
\textit{pairing} interaction ($G_{ij}=1$) corresponds to the
unconventional $\grpso[-]{n+1}$, or $\overline{\grpso{n+1}}$, phase
choice for the \textit{multipole} interaction, and \textit{vice
versa}.

Finally, let us consider how the general $\grpu{n+1}$
scheme~(\ref{eqnchainun}) specializes to the Schwinger realization of
the Lipkin model ($L\<=0$).  With the usual Schwinger angular momentum
operator definitions $J_+\<=b^\dagger s$, $J_-\<=s^\dagger b$, and
$J_z\<=\tfrac12(b^\dagger b-s^\dagger s)$, the subalgebra structure is
\begin{equation}
\label{eqnchainu2}
\renewcommand{\arraystretch}{1.3}
\gqn{\grpu{2}}{}
\supset
\gqn{\grpsu{2}}{J=\tfrac12N}
\supset
\left\lbrace
\begin{array}{c}
\gqn{\grpso[+]{2}\equiv\grpso[x]{2}}{J_x}\\
\gqn{\grpu{1}\equiv\grpso[z]{2}}{J_z=\frac12(N_b-N_s)}\\
\gqn{\grpso[-]{2}\equiv\grpso[y]{2}}{J_y} .
\end{array}
\right.
\end{equation}
The breaking of $\grpu{n}$ into $\grpso[+]{n+1}$, $\grpso[-]{n+1}$,
and $\grpu{n}$ subalgebras therefore has a particularly simple
interpretation, as the projection of the $\grpsu{2}$ angular momentum
alternatively along the $x$, $y$, or $z$ axes.

Furthermore, both the $s$ and $b$ bosonic levels are
singlet levels in the Lipkin model.  It follows, as already noted
for the $s$ boson in~(\ref{eqnchainsu11}), that the
seniorities $v_s$ and $v_b$ are restricted to the values~$0$ or~$1$, with
$v_s\<\cong N_s\mod{2}$ and $v_b\<\cong N_b\mod{2}$.  Thus, the grading quantum
number $g$ defined in Sec.~\ref{secobseigen}, which determines the
Lipkin model parity $\pi\<=(-)^g$, is simply the $\grpsu[b]{1,1}$
quasispin seniority $v_b$.  This formally explains the similarity,
discussed in Sec.~\ref{secobseigen}, between the role of $g$ in the
Lipkin model and that of the $\grpso{n}$ angular momentum in the
higher-dimensional $s$-$b$ algebras [Fig.~\ref{fig-evoln-models}(a,b)].
However, for the Lipkin model ($n\<=1$), there is no $\grpso{n}$
angular momentum dual to the $\grpsu[b]{1,1}$ seniority.

%%%%%%%%%%%%%%%%%%%%%%%%%%%%%%%%%%%%%%%%%%%%%%%%%%%%%%%%%%%%%%%%
%%%%%%%%%%%%%%%%%%%%%%%%%%%%%%%%%%%%%%%%%%%%%%%%%%%%%%%%%%%%%%%%

\section{The Lambert $W$ function}
\label{secapplambert}

The Lambert $W$ function~\cite{corless1996:lambert-w} is implicitly
defined as the solution $y\<=W(x)$ to the equation
\begin{equation}
\label{eqnWdef}
x=y e^y.
\end{equation}
In this appendix, we summarize the essential properties of the $W$
function needed for the present analysis (Sec.~\ref{secclass}).  The
complex analysis, asymptotics, series expansion, \textit{etc.}, of the
$W$ function are considered in detail in
Ref.~\cite{corless1996:lambert-w}.

\begin{figure}
\begin{center}
\includegraphics*[width=0.5\hsize]{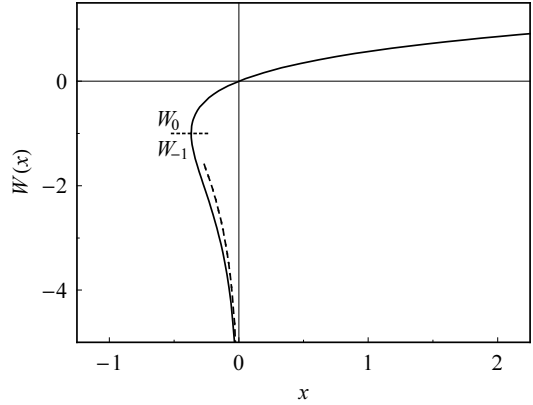}
\end{center}
\caption{The Lambert $W$ function.  For
$-1/e\<<x\<<0$, $W(x)$ is double-valued, with branches
$W_0(x)\<\geq-1$ and $W_{-1}(x)\<\leq-1$.  The asymptotic form
of
$W_{-1}(x)$ given by~(\ref{eqnWasymp}) is
shown for comparison (dashed line).
 }
\label{fig-lambertw}
\end{figure}

Considered as a real-valued function of a real variable, $W(x)$ is
single-valued for $x\<\geq0$ but double-valued for $-1/e\<<x\<<0$,
with branches $W_0(x)\<\geq-1$ and $W_{-1}(x)\<\leq-1$.  These
branches are plotted in Fig.~\ref{fig-lambertw}.  The function has the
asymptotic form~\cite{corless1996:lambert-w,debruijn1961:asymptotic}
\begin{equation}
\label{eqnWasymp}
W_{-1}(x)\sim \log (-x) - \log [-\log(-x)]
\end{equation}
as $x\<\rightarrow0^-$, shown as the dashed curve in Fig.~\ref{fig-lambertw}.

From the defining equation~(\ref{eqnWdef}), it follows that $W$ obeys
the identity
\begin{equation}
\label{eqnWidentity}
\log W(x) = \log x - W(x).
\end{equation}
Differentiation yields
\begin{equation}
\label{eqnWderiv}
W'(x) = \frac{W(x)}{x[1+W(x)]}.
\end{equation}
It also follows from~(\ref{eqnWdef}) that the equation
\begin{equation}
\label{eqnWylogy}
y \log y + c y = x
\end{equation}
has solution $y\<= x / W(e^c x)$, as needed for
Sec.~\ref{secclassasymp}.

%%%%%%%%%%%%%%%%%%%%%%%%%%%%%%%%%%%%%%%%%%%%%%%%%%%%%%%%%%%%%%%%
%%%%%%%%%%%%%%%%%%%%%%%%%%%%%%%%%%%%%%%%%%%%%%%%%%%%%%%%%%%%%%%%

%bibliographystyle{apsrevm}
\input{esqpt.bbl}%bibliography{apsrevm_elsevier,master,theory,expt,data,misc,mc,books,proc,esqpt_special}

\end{document}

%% file: esqpt.bbl
% Bibliography created with apsrevm.bst
\providecommand{\ELSEVIER}{}
\ELSEVIER\newcommand{\identity}[1]{{#1}}